\documentclass{article}
\usepackage[space]{cite}
\usepackage[final]{neurips_2020}

\usepackage{pdflscape}
\usepackage{rotating}

\usepackage[utf8]{inputenc} 
\usepackage[T1]{fontenc}    
\usepackage{hyperref}       
\usepackage{url}            
\usepackage{booktabs}       
\usepackage{amsfonts}       
\usepackage{nicefrac}       
\usepackage{microtype}      
\usepackage{hyperref}
\usepackage{bbm}
\usepackage{mathtools}      
\usepackage{import}
\usepackage{caption} 
\usepackage{wrapfig}
\usepackage{tabularx}
\usepackage{multirow}

\usepackage[space]{cite}
\setcitestyle{square}
\setcitestyle{numbers}
\usepackage{amsmath}
\usepackage{amssymb}
\usepackage{amsthm}
\usepackage{microtype}
\usepackage{graphicx}
\usepackage{subfigure}
\usepackage{dsfont}
\usepackage{booktabs} 
\usepackage{adjustbox}
\usepackage{diagbox} 
\usepackage{etoolbox}
\usepackage{scalerel,stackengine}
\usepackage{bbm}
\usepackage{algorithm}
\usepackage{algpseudocode}

\newtheorem{theorem}{Theorem}
\newtheorem{proposition}{Proposition}
\usepackage{listings}
\usepackage{xparse}
\usepackage{xcolor}
\usepackage{epstopdf}
\usepackage{epsfig}
\usepackage{enumitem}

\usepackage{tikz}
\tikzset{
  treenode/.style = {shape=rectangle, draw, align=center},
  root/.style     = {treenode},
  env/.style      = {treenode},
  dummy/.style    = {circle,draw}
}

\renewcommand{\P}{\mathbb{P}}

\stackMath
\newcommand\autowidehat[1]{%
\savestack{\tmpbox}{\stretchto{%
  \scaleto{%
    \scalerel*[\widthof{\ensuremath{#1}}]{\kern0pt\bigwedge\kern0pt}%
    {\rule[-\textheight/2]{1ex}{\textheight}}
  }{\textheight}%
}{0.5ex}}%
\stackon[1pt]{#1}{\tmpbox}%
}

\newenvironment{talign*}
 {\csname align*\endcsname}
 {\endalign}
\newenvironment{talign}
{\align}
{\endalign}

\setlist[itemize]{leftmargin=5mm,itemsep=0.5mm}
\setlist[enumerate]{leftmargin=*,itemsep=0.5mm}

\title{Bayesian Probabilistic Numerical Integration \\ with Tree-Based Models}

\author{%
  Harrison Zhu, Xing Liu \\
  Imperial College London\\
  \texttt{\{hbz15,xl6116\}@ic.ac.uk} \\
  \And
    Ruya Kang \\
  Brown University \\
  \texttt{ruya$\_$kang@brown.edu} \\
  \And
   Zhichao Shen \\
  University of Oxford \\
  \texttt{zhichao.shen@new.ox.ac.uk} \\
  \AND
  Seth Flaxman \\
  Imperial College London \\
  \texttt{s.flaxman@imperial.ac.uk}\\
  \And
  Fran\c{c}ois-Xavier Briol \\
  University College London \\
  \texttt{f.briol@ucl.ac.uk} \\
}

\begin{document}

\maketitle

\begin{abstract}
Bayesian quadrature (BQ) is a method for solving numerical integration problems in a Bayesian manner, which allows users to quantify their uncertainty about the solution.
The standard approach to BQ is based on a Gaussian process (GP) approximation of the integrand. As a result, BQ is inherently limited to cases where GP approximations can be done in an efficient manner, thus often prohibiting very high-dimensional or non-smooth target functions. 
This paper proposes to tackle this issue with a new Bayesian numerical integration algorithm based on Bayesian Additive Regression Trees (BART) priors, which we call BART-Int.
BART priors are easy to tune and well-suited for discontinuous functions.
We demonstrate that they also lend themselves naturally to a sequential design setting and that explicit convergence rates can be obtained in a variety of settings. 
The advantages and disadvantages of this new methodology are highlighted on a set of benchmark tests including the Genz functions, and on a Bayesian survey design problem. 
\end{abstract}


\section{Introduction}
\label{Introduction}

Numerical integration algorithms are key tools for modern statistics and machine learning. They allow us to tackle many of the integrals commonly arising in those fields and which cannot be computed in closed form. Examples include the computation of posterior expectations in Bayesian statistics, the marginalisation or conditioning of random variables, the computation of normalisation constants, calculations for the EM algorithm, or even the approximation of solutions to differential equations. 

This paper will focus on the task of approximating the integral of some function $f:\mathcal{X} \rightarrow \mathbb{R}$ which is integrable with respect to some distribution $\Pi$ (whose density with respect to the Lebesgue measure is denoted $\pi$) over some domain $\mathcal{X}\subseteq \mathbb{R}^d$ ($d\in \mathbb{N}_+ = \mathbb{N}\backslash \{ 0\}$):
\begin{talign}
	\Pi[f] :=\int_{\mathcal{X}} f(x) d\Pi(x) = \int_{\mathcal{X}} f(x) \pi(x) dx.
\label{eq:integral}
\end{talign}
A large number of methods have been developed to tackle this problem.
Classical quadrature rules \cite{Davis2007} tend to be limited to low-dimensional problems and integrals against a small class of probability measures. Alternatively, Monte Carlo integration (MI) \cite{Robert2004} only requires sampling independently from $\Pi$ and enjoys a convergence rate of $\mathcal{O}(n^{-1/2})$. When $\pi$ is unnormalised, sequential Monte Carlo (SMC) \cite{Moral2006} or Markov chain Monte Carlo (MCMC) \cite{Robert2004} samplers can be used and enjoy a similar rate.  Quasi-Monte Carlo (QMC) \cite{Dick2013} can improve on this convergence rate, but is limited to integration against a uniform measure in a hyper-cube (or simple transformations of this problem). We note that all of these methods are \emph{quadrature rules}: they take the form $\hat{\Pi}[f]:=\sum_{i=1}^n w_i f(x_i)$ where $\{x_i \}_{i=1}^n\subset \mathcal{X}$ are design points and $\{ w_i\}_{i=1}^n\subset \mathbb{R}$ are weights. Finally, the Laplace approximation \cite{Rue2016} and variational inference \cite{Blei2017} can be used, but these cannot be guaranteed to provide accurate estimates as $n$ grows.

Although all of the methods above are commonly used in practice, they all lack a straightforward non-asymptotic approach to quantifying our uncertainty about the value of $\Pi[f]$ after a finite number of function evaluations. This is not necessarily a problem for applications where $n$ can be taken to be large relative to the difficulty of the problem. However, those methods will be sub-optimal when $f$ is both difficult to approximate and expensive to evaluate since the asymptotic results will not hold.

An alternative approach comes from the field of probabilistic numerics \cite{Suldin1959,Larkin1972,Diaconis1988,OHagan1992,Hennig2015}, and in particular Bayesian probabilistic numerical methods \cite{Cockayne2017BPNM}, which frame problems in numerical analysis as statistical estimation tasks. This allows for the quantification of uncertainty surrounding the value of a quantity of interest using probabilistic statements valid for finite $n$. For integration, the main approach is called Bayesian quadrature (BQ) \cite{OHagan1991,Diaconis1988,Rasmussen:2002:BMC:2968618.2968681,Briol2019PI}. Its name comes from the fact that the estimator takes the form of a quadrature rule, but also carries a Bayesian interpretation. However, we will see that not all Bayesian estimators have this property. For this reason, we prefer the general name of \textit{Bayesian probabilistic numerical integration (BPNI)}.

BPNI starts by positing a prior distribution for the integrand $f$. It then computes the posterior distribution given values of $f$ at some design points, and finally considers the implied distribution on $\Pi[f]$. The main advantage of BPNI is that all our knowledge about the problem can be straightforwardly implemented in a prior, and the posterior distribution on $\Pi[f]$ can allow us to quantify our uncertainty about the exact value of this quantity. BPNI critically relies upon a flexible statistical model for the integrand $f$, which is almost always chosen to be a Gaussian process (GP). In this case, we have a quadrature rule which we will call GP-BQ. GPs have the convenient property that the posterior distribution can be obtained in closed form for interpolation and regression with Gaussian noise. They also have well-studied concentration rates which carry over to the corresponding GP-BQ methods \cite{Briol2019PI,Kanagawa2017,Kanagawa2019,Wynne2020}.

GP-BQ has received a lot of attention in recent years. Efficient deterministic \cite{Karvonen2017symmetric,Karvonen2019},  adaptive \cite{Gunter2014,briol2015frank,Fisher2020,Jiang2019} and randomised \cite{belhadji2019kernel} point-selection schemes have been designed. The method was extended to cases where $\pi$ is not available \cite{Oates2017heart}, and when multiple integrals are computed simultaneously \cite{Xi2018MultiOutput,Gessner2019}. It has been applied to fields ranging from econometrics \cite{Oettershagen2017} to computer graphics \cite{Brouillat2009} and robotics \cite{Quann2018}. While GPs are virtually the only choice that has been explored in the BPNI literature, in practice they suffer from a number of challenges. These include:

\begin{enumerate}

    \item \textbf{Discontinuities:} GPs are not suited for integration problems where the integrand is discontinuous -- a common challenge in applications \cite{Cornford1999,Mohammadi2019}. As a result, they may lead to poor predictions and unreliable uncertainty estimates for these problems.

    \item \textbf{Computational cost:} A major issue with GPs is their cubic cost in the number of data points. This can be mitigated using approximate GPs, but these do not necessarily lead to closed-form estimators. Exceptions include the work of \cite{Karvonen2017symmetric} and \cite{Jagadeeswaran2018} which allow for near-linear computational cost, but require $\Pi$ and $\{x_i\}_{i=1}^n$ to exhibit symmetry properties which may not hold in practice.

    \item \textbf{High dimensions:} Few applications of GP-BQ exists where $d>10$ due to the curse of dimensionality. This issue is closely linked to the computational cost issue since the number of points needed to approximate functions well will grow exponentially with dimension. However, it is sometimes possible to use sparsity of $f$ to lower this cost \cite{Briol2019PI}.
    
\end{enumerate}

To tackle some of these issues, we propose to use an alternative model based on trees. The specific model is called Bayesian additive regression trees (BART) \cite{BART,Hill2020}, and as a result we call this BPNI method \textit{BART integration (BART-Int)}. BART is a sum-of-trees model similar in spirit and effectiveness to random forests. It has been successfully deployed in a number of settings including for causal inference \cite{Hill2011, dorie2019automated}, genomics \cite{Ding2012}, behavioural sciences \cite{yeager2019national} and Bayesian optimisation \cite{Chipman2012}. Extensive experimental results in the literature have shown that BART does not usually overfit, thanks to its ensemble structure, and that it only requires a limited amount of hyperparameter tuning. 

\textit{\textbf{Contributions of the paper:}}
This paper derives a novel BPNI algorithm based on Bayesian priors with a tree structure. We show that BART is a natural choice for BPNI which can be preferable to GPs in any of the three settings highlighted. This is done through a mix of theory and experiments. On the theoretical side, we prove asymptotic convergence rates for BPNI with MCMC approximations. This result may be of independent interest since it is applicable for any nonparametric method for which a concentration rate is known. On the experimental side, we compare our method to GP-BQ on a range of problems including the Genz test functions, a rare-event simulation problem and a Bayesian survey design problem.

\section{Background: Bayesian Regression with GPs and BART}
\label{background}

\label{bart_functino}

Before presenting BART-Int, this section introduces background material on Bayesian regression with GPs and BART. Given data $(X,y)$ consisting of design points $X = (x_1,\ldots,x_n)^\top \subset \mathcal{X}^n$ and responses $y = (y_1,\ldots,y_n)^\top \subset \mathbb{R}^n$, the regression problem is to recover an unknown function $f:\mathcal{X}\rightarrow\mathbb{R}$ for which we have access to noisy measurements $y$. These are usually assumed to be independently and identically distributed (i.i.d.\@) Gaussian: $y_i = f(x_i) + \epsilon_i$, $\epsilon_i \sim \mathcal{N}(0,\sigma^{2})$ for $i=1,\ldots,n$, where $\mathcal{N}(a,b)$ denotes a Gaussian distribution with mean $a$ and variance $b$, and $N(\cdot |a, b)$ will denote its density.

Bayesian (nonparametric) regression \cite{Ghosal2017} consists of specifying a (nonparametric) prior distribution over the function $f$ and conditioning on the observations  to obtain a posterior distribution on $f$. This prior is often chosen to be some (real-valued) stochastic process $g:\mathcal{X} \times \Omega \rightarrow \mathbb{R}$ where $(\Omega,\Sigma,\mathbb{P})$ is some probability space. These can be thought of as a random function since $\forall \omega \in \Omega$, $g(\cdot;\omega)$ will be a function, and $\forall x \in \mathcal{X}$, $g(x;\cdot)$ will be a random variable. 

A common choice of prior is a GP \cite{Rasmussen2006}. A GP is fully determined by its mean function $\mu:\mathcal{X}\rightarrow \mathbb{R}$ and covariance function (or kernel) $k:\mathcal{X}\times \mathcal{X}\rightarrow \mathbb{R}$, and hence often denoted by $\mathcal{GP}(\mu, k)$. Due to conjugacy properties of the Gaussian distribution, the posterior $f$ after conditioning on $(X,y)$ is once again a GP with posterior mean $\tilde{\mu}(x) = \mu(x) + k_{x,X} (k_{X,X}+\sigma^2 I)^{-1}(y - \mu_X)$ and covariance $\tilde{k}(x, x') = k(x,x') - k_{x, X} (k_{X,X}+\sigma^2 I)^{-1} k_{X, x'}$, where $k_{X,X}\coloneqq (k(x_i, x_j))_{i,j=1}^n$, $k_{x, X}\coloneqq (k(x,x_1), \ldots, k(x, x_n))$, $k_{X,x}=k_{x,X}^\top$, $\mu_X \coloneqq (\mu(x_1),\ldots,\mu(x_n))^\top$ and $I$ is the $n \times n$ identity matrix. The properties of GP posteriors are inherited from $\mu$ and $k$. Whenever these are continuous, the posterior mean will also be continuous. This is the case for the most common choices of kernels, such as the Gaussian or Mat\'ern kernels with sufficient smoothness. In those cases, discontinuities would have to be inputted manually or inferred from data \cite{Cornford1999}, making the modelling of functions with a large number of discontinuities a challenge for most GP models.

In this paper, we consider instead models based on tree structures. Popular examples include Bayesian CART \cite{Denison1998}, dynamic regression trees \cite{Taddy2011} and Mondrian forests \cite{Lakshminarayanan2016}. We will focus on BART \cite{BART,Hill2020} due to its strong empirical performance and theoretical results. BART is a model which consists of a combination of regression trees. A \textit{regression tree} is any step function $g_{\mathcal{T}, \beta}:\mathcal{X}\rightarrow\mathbb{R}$ with $K$ leaves/partitions: $g_{\mathcal{T}, \beta}(x)\coloneqq  \sum_{k=1}^K \beta_{k}\mathbbm{1}_{\chi_k}(x)$. Here, we denote by $\mathcal{T}\coloneqq\{ \chi_k\}_{k=1}^K$, where $\chi_k \subset \mathcal{X}$, the partition of the domain, and by ${\beta}\coloneqq(\beta_1,\ldots, \beta_K)^\top \in \mathbb{R}^K$ the leaf values. The function $\mathbbm{1}_{\chi_k}:\mathcal{X}\rightarrow \mathbb{R}$ is an indicator function taking value $1$ whenever $x \in \chi_k$ and $0$ otherwise. 

Similarly to decision trees and random forest, a single tree comprising of a large number of leafs might overfit to the training data. As a result, it is common to use an ensemble of shallow trees. We will call $T$-\textit{additive regression tree} any function which takes the form of a sum of regression trees: $g_{\mathcal{E}, \mathcal{B}}(x)\coloneqq \sum_{t=1}^T g_{\mathcal{T}_t, {\beta}_t}(x),$ where $\mathcal{E}\coloneqq\{\mathcal{T}_t \}_{t=1}^T$ and $\mathcal{B}\coloneqq \{\beta_t\}_{t=1}^T$. Finally, we call \textit{Bayesian additive regression tree} (BART) any distribution on the family of $T$-additive regression trees. Such a distribution can be constructed by specifying a (prior) distribution on the
partition $\mathcal{E}$ and leaf values $\mathcal{B}$. BART is hence a stochastic process whose sample space $\Omega$ consists of the product space of $K_t$-partitions of $\mathcal{X}$ and $\mathbb{R}^{K_t}$. For simplicity, BART is usually restricted to an approximation domain $\mathcal{X} = [0,1]^d$, but this is not a requirement in full generality. 

Denote by $p$ the density of this prior distribution. We will follow the majority of the BART literature \cite{CART, BART, rockova2019theory} and use prior models which factorise in the following way: $p(\mathcal{E}, \mathcal{B}) \coloneqq \prod_{t=1}^T p(\mathcal{T}_t)p(\beta_t|\mathcal{T}_t)$ and $ p(\beta_t|\mathcal{T}_t) \coloneqq \prod_{k=1}^{K_t} N(\beta_{t,k}|0, 1/(16T))$, where $T$ is the number of trees and $K_t$ is the number of leaves in the $t^{\text{th}}$ tree. The construction of the distribution on partitions is usually itself done via a tree generating stochastic process; see further details in \cite{BART} (and a full description in Appendix \ref{appendix:BART}).

Given this prior $\mathbb{P}$ on $T$-additive trees, we can condition on $(X,y)$ to obtain a posterior $\mathbb{P}_n$ (with density $p_n$). We focus on regression with i.i.d.\@ Gaussian noise, but several generalisations exist \cite{pratola2017}. The corresponding posterior distribution on parameters will then imply a posterior distribution on $T$-additive trees. BART is hence another Bayesian model which we can use to approximate the integrand, using for example the posterior mean: $g^n(x) = \mathbb{E}_{\mathbb{P}_n}[f(x)] = \mathbb{E}[f(x)|X,y]  =  \int_{\Omega} g_{\mathcal{E},\mathcal{B}}(x) p_n(\mathcal{E},\mathcal{B}) d\mathcal{E} d \mathcal{B}$. 

Unfortunately, this is not available in closed form and needs to be approximated. This is usually done using Markov chain Monte Carlo (MCMC) methods \cite{BART,Pratola2016} (although optimisation methods can also be used \cite{he2019xbart}). At each iteration $j$, the MCMC algorithm produces $T$ regression trees. More precisely, for the $t^{\text{th}}$ regression tree, the algorithm returns a set of leaf values $\textstyle \{\beta_{t,k}^j\}_{k=1}^{K_{t,j}} $ and a partition of the domain $\textstyle \{\chi_{t,k}^j\}_{k=1}^{K_{t,j}}$ where $K_{t,j}$ is the number of leaves for tree $t$ at iteration $j$. This gives a T-additive regression tree which we will denote $g^n_j$. After $m$ iterations, the MCMC mean is given by:
\begin{talign}
    \hat{g}^n(x) & = \frac{1}{m}  \sum_{j=1}^m g^n_j(x)  = \frac{1}{m}  \sum_{j=1}^m \sum_{t=1}^T \sum_{k=1}^{K_{t,j}} \beta_{t,k}^j \mathbbm{1}_{\chi^j_{t, k}}(x). \label{eq:MCMCapprox_f}
\end{talign}
A similar expression can be obtained for the variance of our posterior on $f$ at $x \in \mathcal{X}$. However, when compared to GPs, the entire BART posterior can in fact provide us with a more complex, possibly multimodal, distribution on $f$.


\section{Numerical Integration with Bayesian Additive Regression Trees}
\label{BART}

The Bayesian approach to numerical integration can be succinctly summarised in the following steps:
\begin{enumerate}
    \item \emph{Posit a Bayesian prior on the integrand $f$}. This consists of a stochastic process $g:\mathcal{X}\times \Omega \rightarrow \mathbb{R}$ together with a prior measure $\mathbb{P}$ on $\Omega$.
    \item \emph{Condition this prior on the data $(X,y)$ to get a posterior on $f$}. This is given in terms of the posterior measure $\mathbb{P}_n$, and the posterior mean is $g^n := \mathbb{E}_{\mathbb{P}_n}[ g] = \mathbb{E}[g|X,y]$.
    \item \emph{Obtain the posterior distribution on $\Pi[f]$}. This is given by the marginal distribution obtained by integrating out $\Pi$, and a point estimate for the posterior mean $\Pi[f]$, given by $\hat{\Pi}_{\text{BPNI}}[f] = \Pi[g^n]$. 
\end{enumerate}
We remark that the posterior distribution on $\Pi[f]$, or even just the estimate $\hat{\Pi}_{\text{BPNI}}[f]$, may not be available in closed-form. In that case, MCMC estimates can be used (see the following section). In the case of GP-BQ with a prior $\mathcal{GP}(\mu,k)$, the posterior on $\Pi[f]$ is available in closed-form \cite{Briol2019PI} and is given by a Gaussian with mean $\mathbb{E}[\Pi [f] |X,y] = \Pi[k_{\cdot,X}] (k_{X,X}+\sigma^2I)^{-1}(y-\mu_X)$ and variance $\mathbb{V}[\Pi [f] | X,y]  = \Pi\bar{\Pi}[k]- \Pi[k_{\cdot,X}](k_{X,X}+\sigma^2 I)^{-1}\Pi[k_{X,\cdot}],$ with $\Pi[k_{\cdot,X}] = (\Pi[k(\cdot,x_1)],\ldots,\Pi[k(\cdot,x_n)])$, $\Pi[k_{X,\cdot}] = \Pi[k_{\cdot,X}]^\top$ and $\bar{\Pi}$ denotes the integration functional with respect to the second input. The posterior mean can be expressed as a quadrature rule whose weights depend on $X$, $k$ and $\sigma$. This expression will be available in closed form whenever $\Pi[k(\cdot,x)]$, known as the \textit{kernel mean}, is available in closed form. This is potentially a restrictive condition, but these integrals can themselves be approximated using quadrature, or the kernels could be constructed with this condition in mind  \cite{Oates2017,Oates2016CF2}.

A significant challenge for GP-BQ is the computational cost, which is $\mathcal{O}(n^3)$ due to the need to invert an $n \times n$ matrix. This can possibly be alleviated using fast GP algorithms, but often at the cost of introducing approximations to the posterior. The GP-BQ method is hence more expensive than most standard MC methods, but this should be balanced with significantly faster convergence rates; see \cite{Briol2019PI,Kanagawa2019,Wynne2020}. Overall, this method should hence be preferred to MC methods whenever the number of quadrature points $n$ is small or moderate, or when the integrand is itself expensive.

We now derive a new BPNI method via BART. Suppose that $g$, $\mathbb{P}_0$ are a BART prior (as described in Sec.~\ref{bart_functino}) and we have data $(X,y)$; the posterior on $f$ implies a posterior on $\Pi[f]$. An important distinction with GP-BQ is that the posterior on $\Pi[f]$ is not available in closed-form, and must be approximated. It is also not fully specified by the mean and variance, and may, in fact, be multi-modal. However, the mean of this posterior can be useful if a point estimate for $\Pi[f]$ is required, and the variance can be a useful summary of uncertainty. Expressions for both of these quantities are included in the proposition below for completeness.
\vspace{2mm}
\begin{proposition}[\textbf{MCMC for BART-Int}]\label{eq:BARTBQ_mean&var}
The MCMC approximation of $\Pi[f]$ consists of samples:
\begin{talign*}
s_j = \Pi\left[g^n_j\right] = \sum_{t=1}^T\sum_{k=1}^{K_{t,j}} \beta^j_{t, k}\Pi\left[\mathbbm{1}_{\chi^j_{t, k}}\right], \qquad j\in\{1,\ldots,m\}.
\end{talign*}
These lead to estimates of the mean $ \textstyle \widehat{\mathbb{E}}_{\text{m}}[\Pi[f]|X,y] = \Pi[\hat{g}^n] =\frac{1}{m}\sum_{j=1}^m s_j$ and variance $\textstyle \widehat{\mathbb{V}}_{\text{m}}[\Pi[f]|X,y]  = \frac{1}{(m-1)} \sum_{j=1}^m (s_j - \Pi[\hat{g}^n])^2$.
\label{prop:mcmc}
\end{proposition}
The posterior samples $\{s_j\}_{j=1}^m$ can only be obtained in a closed form whenever probabilities of the form $\Pi(\chi)$ can be computed for any $\chi$ in the partition. This issue corresponds to the well-known issue of intractable kernel means for GP-BQ. The simplest case for which $\Pi(\chi)$ can be computed is when $\Pi$ is the uniform distribution. If $\pi$ is available in closed form, one option is to model $f \pi$ as the integrand, in which case the integral is once again against a uniform. However, this makes the specification of a BART prior challenging since very little will usually be known about $f \pi$, and as a result the uncertainty quantification may be unreliable. 

Another solution, used in this paper and commonly used for intractable kernel means, is to approximate these probabilities using samples $\{\tilde{x}_i\}_{i=1}^l$ representative of $\Pi$. This leads to approximate MCMC samples and an estimate of $\Pi[f]$ of the form:
\begin{talign*}
&\hat{s}^l_j=\frac{1}{l}\sum_{i=1}^{l} g^n_j(\tilde{x}_i)= \frac{1}{l}\sum_{i=1}^{l} \sum_{t=1}^T\sum_{k=1}^{K_{t,j}} \beta^j_{t, k}\mathbbm{1}_{\chi^j_{t, k}}(\tilde{x}_i), \\
&\widehat{\mathbb{E}}^l_{\text{m}}[\Pi[f]|X,y] = \frac{1}{m}\sum_{j=1}^m \hat{s}_j^l.  \qquad  \widehat{\mathbb{V}}^l_{\text{m}}[\Pi[f]|X,y]  = \frac{1}{(m-1)} \sum_{j=1}^m \left(\hat{s}^l_j - \left( \frac{1}{m}\sum_{k=1}^m \hat{s}_k^l\right)\right)^2.
\end{talign*}
Using this approach might be counter-intuitive since it means that some of the uncertainty is not quantified in a Bayesian manner. However, when the integrand $f$ is expensive but a very large number of data points are available (i.e. $n \ll l$), the additional error can be made to be negligible relative to the overall integration error. These settings are common in practice, as will be shown in Sec. \ref{Performance}.

A significant advantage of BART-Int over GP-BQ is the computational cost. For BART, the cost is $\mathcal{O}(T m n)$ (or $\mathcal{O}(T m (n+l))$), although the constant depends on the properties of the trees (see \cite{Pratola2014} for a detailed analysis). When $n$ is large, this can be much cheaper than the $\mathcal{O}(n^3)$ for GP-BQ. 
We note that $\{x_i\}_{i=1}^n$ can be selected through any quadrature rule. However, when $f$ is computationally expensive, it may be preferable to adaptively select $\{x_i\}_{i=1}^n$ using tools from experimental design and active learning \cite{Santner2018}. Recent advances in this direction in numerical integration focus on minimising some distance between $\Pi$ and $\{x_i\}_{i=1}^n$ (see e.g.\@ kernel herding \cite{Chen2010}, Stein Variational Gradient Descent \cite{Liu2016SVGD} and Stein points \cite{Chen2018}). All of these methods could be combined with our BART-Int estimators, but such an approach may be sub-optimal in the sense that our objective is not to approximate $\Pi$, but only $\Pi[f]$. We propose instead a method which focuses on improving the fit of the BART posterior mean. Our approach is hence closer to previous active learning strategies for GP-BQ, including the work of \cite{Gunter2014,briol2015frank}. It is summarised in Algorithm~\ref{alg:SQ}. 

\begin{algorithm}[h!]
  \caption{Sequential Design for BART-Int}
  \label{alg:SQ}
\begin{algorithmic}[1]
  \State {\bfseries Inputs:} Initial point(s) $X$ (set of size $n_{\text{ini}}$), response(s) $y$, number of candidate design points $S$, total number of points $n$, acquisition criterion $J$ and number of MCMC samples $m$.
  \vspace{1mm}
  \For{$i \in \{ n_{\text{ini}},\ldots,n-1\}$ iterations}
  \State Obtain $m$ posterior samples (given $X^{i},y^{i}$) \& pick a candidate set $\mathcal{C} = \{c_1, \ldots, c_S\} \subset \mathcal{X}$.
  \State Find $\,c^* = \mbox{argmax}_{c \in \mathcal{C} } J(c)$ \& set $X^{i+1} \leftarrow X^i \cup \{c^*\}$, $\; y^{i+1}\leftarrow y^i\cup \{ y_c^{*}\}, \; y_{c}^*
  = f(c^*)+\epsilon^*$.
  \EndFor
\end{algorithmic}
\end{algorithm}
The second step consists of comparing the suitability of each point in $\mathcal{C}$ according to some criterion $J:\mathcal{X}\rightarrow \mathbb{R}$, then adding the best point to our design set. At iteration $i$, we propose to use
\begin{talign}
    J(c) =  \widehat{\mathbb{V}}_{m}[f(c)\pi(c)|X^i,y^i],
\end{talign}
which leads to selecting points where the uncertainty in $f$ is highest. Alternatively, one could choose $c$ which minimises the posterior variance: $J(c)=- \widehat{\mathbb{V}}_{m}[\Pi[f]|X^i \cup \{c\},y^i \cup \{y_c\}]$. However, this would require training a different MCMC sampler for each point in $\mathcal{C}$, which would significantly increase the computational cost.


\section{Theoretical Results}

We now introduce novel concentration results. All proofs are given in Appendix \ref{appendix:proofs}. These results hold for BART-Int, but are also of independent interest since they can lead to rates for other BPNI methods. The results will hold for an integrand in a normed subspace $\mathcal{H}$ of $L^2(\Pi):=\{h:\mathcal{X} \rightarrow \mathbb{R} \text{ s.t. } \|h\|_{L^2(\mathcal{X})}:= \int_{\mathcal{X}} h^2(x) \pi(x) dx < \infty\}$. They will depend on the contraction rate of the posterior onto the integrand, as measured in some empirical norm: $\textstyle \|f\|_n = (\frac{1}{n} \sum_{i=1}^n f(x_i)^2)^{1/2}$. Below, we will use $(X^n,y^n)$ to denote a dataset $(X,y)$ of size $n$.

\vspace{2mm}
\begin{theorem}[\textbf{Concentration Bound for BPNI}]\label{thm:convergence_rate}
Suppose we have a normed space $\mathcal{H} \subseteq L^2(\Pi)$ such that the integrand $f \in \mathcal{H}$. Furthermore, suppose that our BPNI posterior is based on a stochastic process $g$ (jointly measurable over $\mathcal{X} \times \Omega$) and a sequence of data $\{(X^n,y^n)\}_{n \in \mathbb{N}}$ such that $\forall \omega \in \Omega, g(\cdot,\omega) \in \mathcal{H}$ and $\exists N \in \mathbb{N}_+$ such that:
\begin{itemize}
\item [A1.]  $\exists \{\varepsilon_n\}_{n \geq N}$ such that $\lim_{n \rightarrow \infty} \mathbb{P}_n[\|f - g\|_n > A_n \varepsilon_n]=0$ for any $A_n \rightarrow \infty$ as $n \rightarrow \infty$.
    
\item [A2.] $\exists \{\gamma_n\}_{n \geq N}$ with $\gamma_n \rightarrow 0$ as $n \rightarrow \infty$ such that $\sup_{\|h\|_{\mathcal{H}} \leq 1} |\frac{1}{n}\sum_{i=1}^n h(x_i) - \Pi[h]| = O(\gamma_n)$. 
\end{itemize}
Then, we have $\lim_{n\rightarrow \infty} \mathbb{P}_n[| \Pi[f] -  \Pi[g]| > C_n \max(\varepsilon_n,\gamma_n)] = 0$ for any $C_n \rightarrow \infty$ as $n \rightarrow \infty$.
\end{theorem}
We now make a number of remarks on the assumptions and statement of the theorem:
\begin{enumerate}
    \item To guarantee concentration of a BPNI posterior, we require two ingredients: (i) the Bayesian posterior should concentrate on $f$ as $n$ grows, and (ii) the sequence of point sets $X^n$ should give a sequence of quadrature rules which can integrate both $f$ and $g^n$ for any $n$ large enough. The rates at which (i) and (ii) occur then control the overall concentration rate of the BPNI posterior. 
    \item The assumptions in this theorem are stated in a way such that the order in $n$ at which concentration of the BPNI posterior occurs is explicit. In particular, A1 is a standard assumption in Bayesian nonparametric and guarantees concentration of the posterior on $f$ at rate controlled by $O(\varepsilon_n)$. Similarly, A2 is a common assumption for quadrature rules and guarantees a quadrature rate of $O(\gamma_n)$. The resulting concentration rate of the BPNI posterior is then controlled by $O(\max(\varepsilon_n,\gamma_n))$, the slowest of these two rates. 
    \item A number of quadrature rules can integrate functions in $L^2(\Pi)$ at a rate $O(n^{-1/2})$ (i.e. $\gamma_n =n^{-1/2}$), but have faster worst-case integration rates for subspaces of $L^2(\Pi)$ which are sufficiently regular. For this reason, we introduced the subspace $\mathcal{H}$ and the conditions that $f \in \mathcal{H}$ and $ \forall \omega \in \Omega$, $ g(\cdot,\omega) \in \mathcal{H}$, in order to allow for faster rates for $\gamma_n$. 
    \item Results of the form in A1 are available for a variety of Bayesian nonparametric models \cite{Ghosal2017}, including GPs \cite{Vaart2011} and Bayesian neural networks \cite{Polson2018}. This is a rapidly evolving field and most results were only derived in recent years. The advantage of Thm. \ref{thm:convergence_rate} is that any contraction result derived in the future can be plugged in to understand the corresponding BPNI method.  
    
    \item Assumption A2 will hold for most reasonable choices of design points. For example, it would hold for $V$-uniformly ergodic MCMC when $\mathcal{H}=\{h \in L^2(\Pi): \sup_{x \in \mathcal{X}} |h(x)|/V(x) < \infty \}$ (under regularity conditions given in \cite{Roberts2004}). Note that Assumption A1 may also depend on the design points, and a discussion of various cases can be found in \cite{jeong_art_2020}.
    
    \item As for BART-Int, we might sometimes need to approximate the BPNI using samples $\{\tilde{x}_i\}_{i=1}^l$. In that case, it is sufficient for $\{\tilde{x}_i\}_{i=1}^l$ to satisfy an assumption of the form of A2 for Thm. \ref{thm:convergence_rate} to hold. In general, it will be possible to take $l$ much larger than $n$, so we could potentially even allow for slower quadrature rates than $O(\gamma_n)$.
\end{enumerate}

Now that we have discussed our result on concentration of BPNI posteriors, we consider implications for BPNI with tree-based models. Most results are presented for the space of $\alpha$-H\"older continuous functions, denoted $\mathcal{H}^{\alpha}$, where $ \|f\|_{\mathcal{H}^\alpha} \coloneqq \sup_{x,y \in \mathcal{X}} |f(x)-f(y)|/\|x-y\|_2^{\alpha} < \infty$ (when $\alpha=1$, we recover Lipschitz functions). Several results are of interest:

\begin{itemize}
    \item \cite{rockova2017posterior,rockova2019theory} showed that Bayesian CART and BART concentrate on targets which are sums of functions in $\mathcal{H}^{\alpha}$ ($0 < \alpha \leq 1)$ that are constant in $d-d_0$ coordinates at a rate $\varepsilon_n = n^{-\frac{\alpha}{2\alpha+d_0}}\log^{\frac{1}{2}}n$. This does not require a-priori knowledge of the coordinates in which the function is constant.
    
    \item \cite{linero2018bayesian} showed that the soft-BART posterior raised to a fractional power concentrates at the rate $\varepsilon_n = n^{-\frac{\alpha}{2\alpha+d_0}}\log^\beta n+\sqrt{n^{-1}d_0 \log d}$ for any $\beta \geq \alpha(d_0+1)/(2\alpha+d_0)$, even when $\alpha>1$. 
\end{itemize}
These results are not directly comparable with the most of the literature for GP-BQ \cite{Briol2019PI,Kanagawa2017,Kanagawa2019} since these consider the interpolation setting (where $y_i = f(x_i)$). The closest result is in \cite{Wynne2020}, which presents a result for regression with i.i.d. Gaussian noise and convergence in expectation. The rate in \cite{Wynne2020} is identical to those above except for for sparse high-dimensional functions where it is slower. 

Recent Bernstein-von Mises results for BART presented in \cite{Rockova2019BvM} (Thm. 5.2) also provide guarantees on the uncertainty quantification provided by the BART-Int posterior distribution on $\Pi[f]$, and show that in the asymptotic limit of $n\rightarrow \infty$, the Bayesian credible intervals will coincide with frequentist confidence intervals. Similar results were obtained for sparse deep networks in \cite{Wang2020}.

Before concluding, we note that any MCMC approximation required to approximate the posterior will also impact the convergence. This is made more precise through the following result.

\vspace{2mm}
\begin{proposition}[\textbf{MCMC Approximation for BPNI}] \label{prop:MCMC_convergence}
Let $\hat{\mathbb{E}}_m[\Pi[g]]$ denote the MCMC approximation of $\mathbb{E}[\Pi[g]]$ where the expectation is with respect to some measure $\mathbb{P}$. Assume $\Pi[g] \in L^2(\mathbb{P})$, where the Markov chain targets $\mathbb{P}$ and is geometrically ergodic and reversible. Then, $\exists \sigma_{\text{MCMC}} >0$ such that 
\begin{talign*}
\sqrt{m} | \mathbb{E}[\Pi[g]]- \hat{\mathbb{E}}_m[\Pi[g]]| \rightarrow \mathcal{N}(0,\sigma^2_{\text{MCMC}})
\end{talign*}
in distribution as $m \rightarrow \infty$.
\end{proposition}
Geometric ergodicity is a well-studied concept in MCMC theory; see Appendix \ref{sec:MCMC_proof}. Combining Thm. \ref{thm:convergence_rate} and Prop. \ref{prop:MCMC_convergence} gives us some intuition as to how to balance the computational cost of increasing $n$ and $m$ in order to obtain the smallest possible approximation error. Unfortunately, to the best of our knowledge, the ergodicity of MCMC samplers commonly used for BART has not yet been studied.


\section{Numerical Experiments}
\label{Performance}

We now illustrate BART-Int and GP-BQ on a range of synthetic problems and a Bayesian survey design problem. We emphasise that our goal is to compare methods which can provide a Bayesian quantification of uncertainty; as a result, it is very much possible that non-Bayesian methods could have better point estimate performance on some of these problems. For BART-Int, we used the default prior settings in \verb!dbarts! \cite{dorie2018dbarts}, whereas for GP-BQ we used a Mat\'ern kernel whose lengthscale was chosen through maximum likelihood. Further details can be found in Appendix \ref{appendix:experiments}.

\begin{table}[h!]
\centering
 \label{table:genz}
 \begin{minipage}[t]{\linewidth}
 \begin{adjustbox}{max width=\textwidth}
 \begin{tabular}[t]{| c | c | | c | c | c | c | c | c | c |}
    \hline
     Setup & Method  & Cont   & Copeak &Disc & Gaussian & Oscil & Prpeak & Step \\
     \hline
     \hline
    $d=1$  & BART-Int  & 1.21 &  \textbf{7.32e-01} & \textbf{6.06e-02} & \textbf{6.80e-01} & \textbf{228} & 2.58& \textbf{7.85e-03} \\
    $n_\text{ini}=20$ & MI  & 
\textbf{ 7.11e-01} & 
8.99e-01 & 
1.55e-01 &
 8.41e-01 & 
 251 & 
 1.28 & 
5.90e-02 \\
     $n_\text{seq}=20$& GP-BQ  &9.38e-01 &  3.41 & 1.61e-01 & 7.42e-01 & 229& \textbf{9.88e-01}& 2.88e-02\\
     \hline
      $d=10$ & BART-Int  & \textbf{9.10e-04} & 43.8  &\textbf{1.31e-02} 
      &3.37e-03 &1.19e-02 &1.77e-02 &\textbf{4.33e-03} \\
     $n_{\text{ini}}=200$ & MI  &   
     2.78e-03  & 
     \textbf{2.12e-01} & 
     1.33e-01&
     1.18e-02 & 
     8.43e-02 & 
     3.70e-03 & 
     2.46e-02 \\
     $n_\text{seq}=200$&  GP-BQ   & 
     1.52e-03  & 
     2.79e-01 & 
     1.84e-01&
     \textbf{3.02e-03} &
     \textbf{2.39e-03} & 
     \textbf{6.59e-04} & 
     8.36e-03 \\
     \hline
    \end{tabular}
    \end{adjustbox}
\caption{Integration of the six Genz test functions and a step function. The values correspond to the mean absolute percentage error (MAPE) over 20 separate runs.}
    \end{minipage}
    \vspace{-6mm}
\end{table}

\begin{wrapfigure}{r}{0.3\linewidth}
    \vspace{-7mm}
    \includegraphics[width=\linewidth,trim=0.5cm 0 0.5cm 0.2cm, clip]{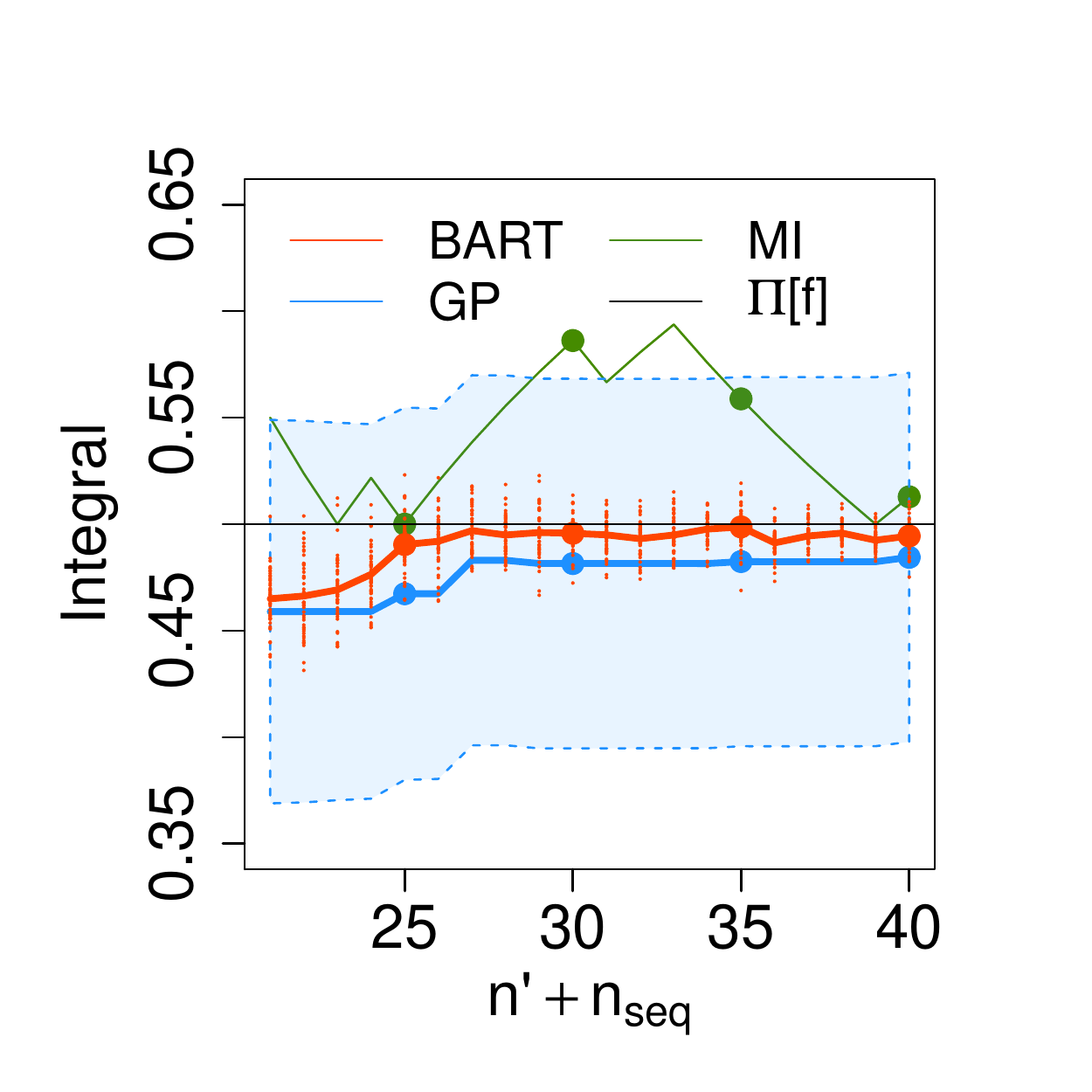}
    \vspace{-5mm}
    \caption{Integration of $f_\text{step}$ against a uniform, $d=1$, $n_{\text{ini}}=20$ and $n_{\text{seq}}=20$. }
    \label{figure:step}
    \vspace{-6mm}
\end{wrapfigure}
\paragraph{Genz Functions} We start with a standard benchmark set of multi-dimensional integrands proposed by Genz \cite{Genz} (see Appendix \ref{appendix:experiments}). This consists of six functions in $\mathcal{X}=[0,1]^d$, where $d$ can be varied to adjust the level of difficulty of the problem. We also add a step function:  $f_\text{step}(x) = \mathbbm{1}_{ \{x_1 \in (0.5, 1] \} }(x)$, where $x_1$ is the first component of $x$. We use $n_{\text{ini}}=20d$ design points, with $n_\text{seq}=20d$ additional points selected by the sequential scheme in Algorithm~\ref{alg:SQ}.

Table~1 shows the mean absolute percentage error (MAPE) of BART-Int ($m=1500$, $T=200$ $m=1000$, $T=50$, with a burn-in of 1000 and keeping every 5 samples afterwards), GP-BQ and Monte Carlo Integration (MI) for all six Genz functions in $d=1$ and $10$. The MAPE is given by given by $\frac{1}{r}\sum_{t=1}^r|\Pi[f]-\Hat{\Pi}_t[ f]|/ |\Pi[f]|$, where $\Hat{\Pi}_t[f]$ for $t=1,\ldots,r$, are estimates of $\Pi[f]$ for $r$ different initial i.i.d. uniform point sets.

BART-Int outperforms GP-BQ when $f$ is not continuous, e.g.\@ the Step functions and the Discontinuous functions. In those cases, the posterior distribution of BART (red dots) is also more concentrated around the truth than GP-BQ (whose 95\% credible intervals are in shaded blue); see Fig \ref{figure:step}. However, GP-BQ is strongest when estimating integrals of smooth continuous in $d=1$ (see all other integrands). This is to be expected since these functions are smooth enough to be well modelled by the GP (see  convergence rates in \cite{Wynne2020}), while the step-function nature of BART makes it more appropriate for non-smooth functions. Finally, in $d=10$, BART-Int outperforms GP-BQ for all integrands as it is adaptive to the important features of the integrand, with the largest gains being once again for discontinuous functions.

\vspace{-3mm}

\begin{wrapfigure}{r}{0.35\linewidth}
 \begin{center} 
 \vspace{-8mm}
    \includegraphics[width=0.8\linewidth,trim=0.6cm 0.7cm 0.7cm 0, clip]{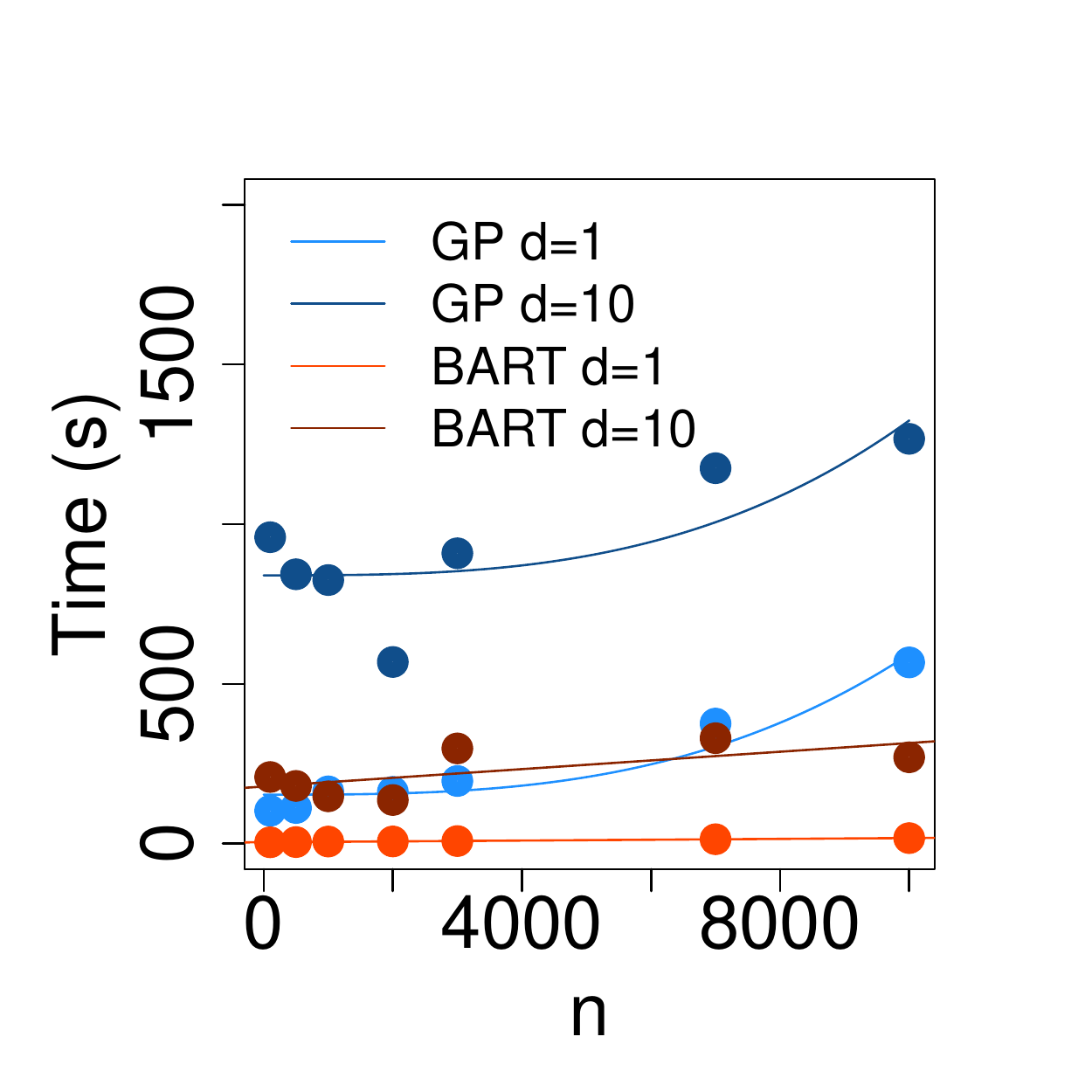}
    \vspace{-1mm}
    \caption{Run-times of GP-BQ \& BART-Int for $\Pi[f_\text{step}]$ (w.o. seq. design). }
    \label{fig:computational_complexity}
    \end{center}
     \vspace{-8mm}
\end{wrapfigure}
\paragraph{Computational Complexity}
The computational complexity of BART-Int is $\mathcal{O}(Tmn)$, which is much cheaper than the $\mathcal{O}(n^3)$ for GP-BQ with even moderately large $n$. This behaviour can be seen empirically in Figure~\ref{fig:computational_complexity}, noting $T$ and $m$ have been fixed. The computational time of BART-Int is based on the assumption that the tree-based operations are constant, which is reasonable so long as the tree sizes are moderate \cite{Pratola2014}. Furthermore, it has been shown empirically that the run time of BART is almost independent of the dimension, $d$, of the data, see e.g. \cite{Pratola2014} and section~6 of \cite{BART}. However, with larger $d$, it would require a longer burn-in period for the parameter space to be fully explored, and thus the run-time for BART (thus also BART-Int) would increase for the purpose of approximation accuracy.

\paragraph{Rare Events Simulation}
Rare-event simulation is an important tool in risk management. It is used in a range of applications, such as for safety testing autonomous driving systems \cite{o2018scalable}, queuing systems \cite{rubino2009rare} and finance \cite{chan2012improved}. Traditionally, rare-event simulation is tackled using Monte Carlo methods \cite{rubino2009rare}, but we propose to use BART-Int instead. 
\begin{wrapfigure}{r}{0.4\linewidth}
\begin{minipage}[t]{\linewidth}
\centering
\begin{adjustbox}{max width=0.99\textwidth} 
\begin{tabular}[t]{| c | c | c | c |}
\hline
  & Method  & MAPE   & Std. Err. \\
 \hline 
  & BART-Int  & 1.71e-01 & 2.56e-02 \\
 $d=5$  & MI  &1.95e-01 & 2.29e-02 \\
  $n=2500$&  GP-BQ  & \textbf{1.68e-01}& 2.09e-02 \\
 \hline
  & BART-Int  &\textbf{1.56e-02} &2.35e-03  \\
 $d=10$  & MI  & 9.98e-01  & 4.47e-04 \\
 $n=5000$ &  GP-BQ   & 2.72e-02& 5.20e-03 \\
 \hline
  & BART-Int & \textbf{8.40e-03}  &1.60e-03 \\
 $d=20$    & MI  &9.94e-01 & 6.34e-04\\
 $n=10000$ &  GP-BQ    &2.92e-02 & 4.90e-03 \\
 \hline
\end{tabular}
\end{adjustbox}
\captionof{table}{Performance on the portfolio loss event over 20 runs.}
\label{tab:high-dimensionality}
\vspace{-5mm}
\end{minipage}
\end{wrapfigure}
Let $\ell:\mathbb{R}^d\rightarrow\mathbb{R}$ be a measurable function representing our loss. The central quantity of interest here is the probability of obtaining a loss larger than some constant $\gamma \in \mathbb{R}$, which can be written as $p_\gamma = \int_{\mathcal{X}} \mathbbm{1}_{\{\ell(x)>\gamma\}} (x)\Pi(dx) $.

We consider a problem of high-dimensional portfolio management adapted from \cite{chan2012improved}. Suppose we have $d$ loans to obligors, each with value $c_i$ for $i=1,\ldots,d$. Let $x_i$ denote the financial strain on loan $i$, and suppose $d_i$ is a thresholds after which default occurs. We assume that the distribution of financial strains is $\text{Exp}(1)$. We can define the portfolio loss as $\ell(x) = \sum_{i=1}^d c_i \mathbbm{1}_{\{x_i > d_i\}} (x)$. Given a threshold $\gamma$, we can compute the probability of making a loss greater than $\gamma$ as $p_\gamma$. 

We set $\gamma=2$, $d_i = 0.5i$ and $c_i=0.2i$. We use the same experimental settings as for the Genz functions, but do not use a sequential design and instead set $n=500d$. The number of post-burn-in samples is chosen to be $10^4$. We can see that for $d=5$ most of the algorithms perform similarly, but as $d$ increases BART-Int gradually performs better, especially when compared to GP-BQ.

\begin{wraptable}{r}{0.3\linewidth}
    \centering
    \vspace{-10mm}
    \includegraphics[width=0.99\linewidth,trim=0.5cm 0.5cm 1cm 0, clip]{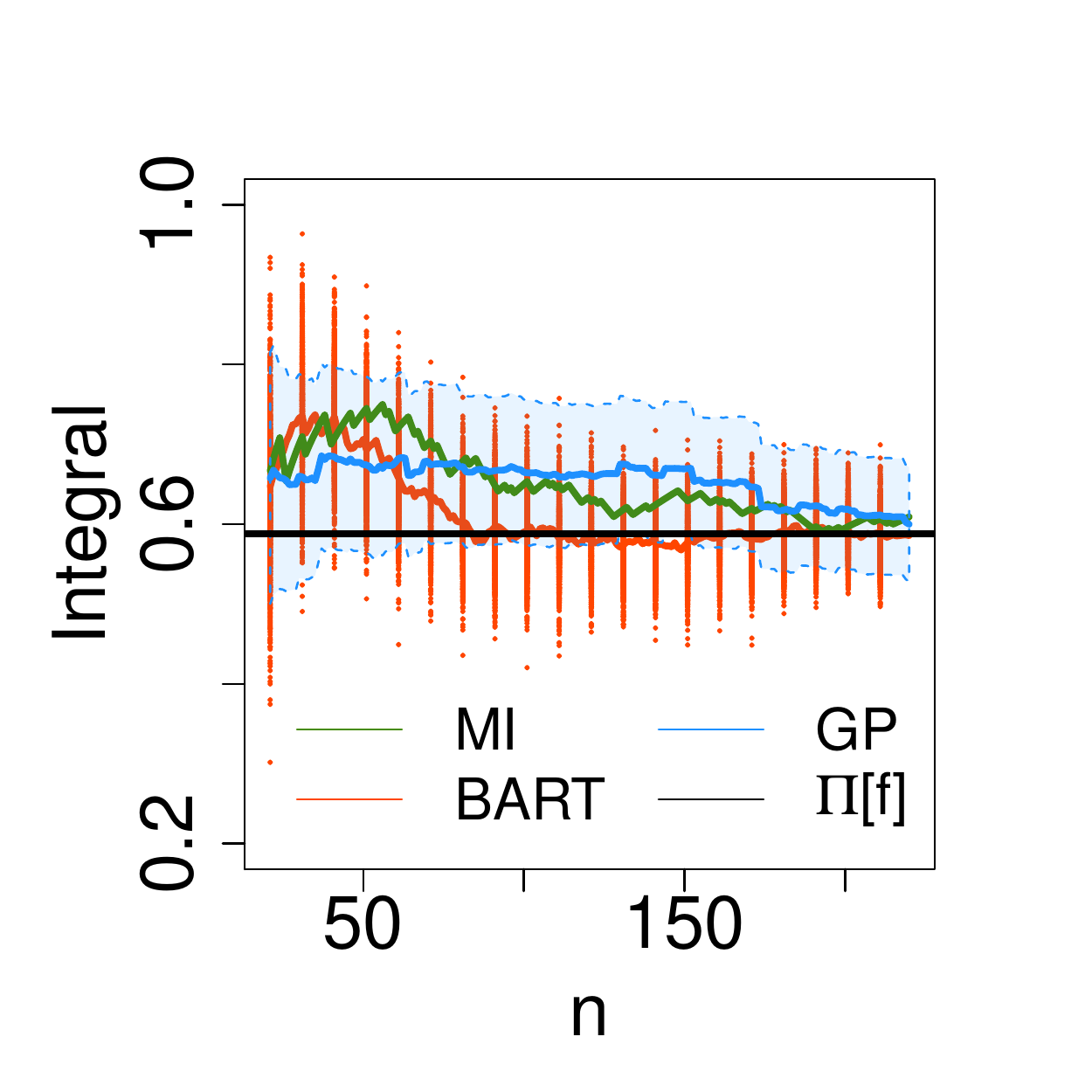}
    \captionof{figure}{Proportion of individuals with log-salary >$10$.}
    \label{fig:survey}
    \vspace{-8mm}
\end{wraptable}
\paragraph{Bayesian Survey Design} In surveys, response variables are collected from a subset of a population based on answers to a set of questions. Survey design concerns sampling strategies to obtain population-representative estimates from these samples. The standard approach is simple or stratified random (Monte Carlo) sampling. Bayesian hierarchical models are also often used in this setting to analyse survey data in order to stabilise estimates and to borrow strength when making sub-population estimates for underrepresented subgroups or locations \cite{gelman2006data}. 

We propose a new approach we term ``Bayesian Survey Design'', using BART-Int to adaptively choose the next individual to survey. 
Assume we have access to a small set of $n_{\text{ini}}$ survey responses with a mixture of continuous and categorical covariates (educational attainment, age, etc).
In addition, we assume there is a much larger set of $S$ individuals for whom demographic variables are known but the response variable is unknown but available for surveying. Such active surveying has previously also been explored in Bayesian active learning \cite{10.5555/3042573.3042683}. To demonstrate this approach, we use individual-level anonymised census data from the United States \cite{us_census_bureau_american_2018} and estimate the proportion of the population whose income is higher than around $\$$22,000, or log-income greater than $10$. This is equivalent to computing the integral of the following indicator function: $f(x) = \mathbbm{1}_{\{\text{log-income} >10 \}}(x)$ where the input consists of all other covariates in the survey and $\Pi$ is the distribution of these covariates in the population. 

\begin{wraptable}{r}{0.4\linewidth}
\vspace{-5mm}
    \begin{adjustbox}{max width=0.9\textwidth} 
    \begin{tabular}[t]{  c | c | c }
        \hline
          Method  & MAPE   & Std. Err. \\
         \hline
         BART-Int  & \textbf{4.66e-02} & 7.38e-03 \\
         MI  &5.68e-02 & 7.01e-03\\
          GP-BQ  & 9.73e-02& 1.44e-02 \\
         \hline
        \end{tabular}
    \end{adjustbox}
    \captionof{table}{Performance on the Survey Design problem over 20 runs.}
    \label{tab:survey_design}
    \vspace{-4mm}
\end{wraptable}

We consider a universe of $454,816$ possible respondents with $8$ demographic covariates. After a one-hot encoding of categorical variables, the dimensionality is $d=24$. We randomly select our initial set (of size $n_\text{ini}=20$) and candidate set (of size $S=10,000$). We compute ground truth using all $454,816$ observations and use Algorithm \ref{alg:SQ} to select $n_\text{seq}=200$ new individuals to survey. As seen in Table~\ref{tab:survey_design} and Figure~\ref{fig:survey}, BART-Int outperforms both MI and GP-BQ. As expected, the posterior distributions contract to the population mean as $n$ increases. Furthermore, the BART-Int posterior (red dots) is centered around the truth, whereas the 95\% credible intervals (blue shaded region) for GP-BQ seem overconfident at the wrong value around $n=100$.


\vspace{-1mm}

\section{Related Work}

Before concluding, we briefly comment on the connections between BART-Int and other tree-based algorithms for integration. Our work is most closely related to GP-BQ, but with tree-based priors. Although trees had not been considered in this literature, the nearest-neighbors method in \cite{llorente2020adaptive} uses GP kernels which depends on indicator functions at Voronoi partitions, and yield approximations to $\Pi[f]$ which closely resembles that in our paper. 

We note that tree-based methods have received significant interest in the literature in numerical integration, but previous work does not come with a Bayesian interpretation. For example, the \texttt{VEGAS} algorithm \cite{PeterLepage1978} from the physics literature can be thought of as a tree-based method to reduce the variance of MI. More recent work also includes MCMC and SMC samplers with proposals estimated using trees \cite{felip2019tree, hanson2011polya}. Finally, \cite{foster2020model} also uses regression trees as surrogates, but again in a non-Bayesian framework.  

\vspace{-2mm} %

\section{Discussion}
\label{Discussion}

\vspace{-2mm}
We proposed a novel BPNI algorithm which uses BART instead of a GP to model the integrand. BART has several advantages in settings where GPs do not perform well. It is easy to tune, robust to overfitting and enjoys a low computational complexity. It is also a natural model for discontinuous functions and can automatically adapt to sparse high-dimensional functions. We demonstrated these advantages through theoretical results, including contraction of the posterior distribution onto the value of the integral, and through a set of benchmark tests and a survey design experiment. However, it is also important to note that BART-Int did not perform as well as GP-BQ in certain low dimensional and smooth settings. We therefore see BART-Int as a useful addition to the toolset of BPNI which complements, rather than replaces, GP-BQ. 

There are a number of potential directions for future research. From an applications-viewpoint, there are interesting parallels to be made with algorithms used to estimate treatment effects in causal inference, and we foresee some possible use for BART-Int in this field. This was discussed in more details in an opinion piece published in \cite{hahn2020}. In terms of theory, little is known about the ergodicity of the MCMC sampler used to sample posterior trees. Further work in this area could help practitioners understand the impact of the MCMC approximation on the estimate of the integral for BART-Int. In terms of methodology, while the sequential design approach proposed in this paper uses the (estimated) posterior variance to select new query points, it would be interesting to explore how the use of other acquisition functions may improve the estimates. Similarly, extensions of BART, such as soft-BART \cite{linero2018bayesian} and heteroscedastic-BART \cite{pratola2017}, should also be of practical interest, as they may provide additional advantages over GP-based methods. Finally, some tree-based models can naturally handle categorical variables without the need for one-hot encoding; this could significantly improve the performance for problems such as the Bayesian survey design experiment in this paper.

\section*{Broader Impact}
Our paper is about numerical integration, a common computational problem in machine learning. It provides an approach to improve the accuracy of such approximations, as well as obtaining credible intervals representing our uncertainty about the value of the integral. The increased accuracy of the method may allow the users to reduce their computational requirements, which may, further down the line, have some impact on mitigating the impact of large computer clusters on climate change. 

However, the broader impact of the method will mostly depend on the applications of the algorithms that it will be used to enhance: applications to ethical algorithms will have a positive impact, but application to unethical algorithms will also have a negative impact. For example, in the Bayesian survey design problem studied in the numerical experiments section, we have shown that there is potential for application in predicting population proportions resulting from binary outcome variables. The categorical variables for instance, however, may be unethically used as explanatory variables for certain problems, and this would be an issue depending on how the conclusions are drawn.

\begin{ack}
The authors would like to thank Matthew Fisher, Takuo Matsubara and Chris Oates for useful feedback on a previous draft of the paper. HZ, XL were supported by the EPSRC Centre for Doctoral Training in Modern Statistics and Statistical Machine Learning (EP/S023151/1) and the Department of Mathematics of Imperial College London. HZ was supported by Cervest Limited. XL was supported by the President’s PhD Scholarships of Imperial College London. RK and ZS were supported by the Undergraduate Research Opportunities Programme of Imperial College London. FXB was supported by the Lloyd's Register Foundation Programme on Data-Centric Engineering and the Alan Turing Institute under the EPSRC grant [EP/N510129/1], and through an Amazon Research Award on ``Transfer Learning for Numerical Integration in Expensive Machine Learning Systems''.
\end{ack}
\bibliographystyle{abbrv}

\begin{thebibliography}{100}

\bibitem{wolframalpha}
Wolfram alpha: computational intelligence.
\newblock Accessed: 2018-01-23.

\bibitem{doi:10.1080/02664760802684177}
G.~Baio and M.~Blangiardo.
\newblock Bayesian hierarchical model for the prediction of football results.
\newblock {\em Journal of Applied Statistics}, 37(2):253--264, 2010.

\bibitem{belhadji2019kernel}
A.~Belhadji, R.~Bardenet, and P.~Chainais.
\newblock Kernel quadrature with {DPP}s.
\newblock In {\em Advances in Neural Information Processing Systems}, pages
  12927--12937, 2019.

\bibitem{Blei2017}
D.~M. Blei, A.~Kucukelbir, and J.~D. McAuliffe.
\newblock {Variational inference: A review for statisticians}.
\newblock {\em Journal of the American Statistical Association},
  112(518):859--877, 2017.

\bibitem{briol2015frank}
F.-X. Briol, C.~Oates, M.~Girolami, and M.~A. Osborne.
\newblock Frank-{W}olfe {B}ayesian quadrature: {P}robabilistic integration with
  theoretical guarantees.
\newblock In {\em Advances in Neural Information Processing Systems}, pages
  1162--1170, 2015.

\bibitem{Briol2019PI}
F.-X. Briol, C.~J. Oates, M.~Girolami, M.~A. Osborne, and D.~Sejdinovic.
\newblock {Probabilistic integration: A role in statistical computation? (with
  discussion)}.
\newblock {\em Statistical Science}, 34(1):1--22, 2019.

\bibitem{Brouillat2009}
J.~Brouillat, C.~Bouville, B.~Loos, C.~Hansen, and K.~Bouatouch.
\newblock {A Bayesian Monte Carlo approach to global illumination}.
\newblock {\em Computer Graphics Forum}, 28(8):2315--2329, 2009.

\bibitem{ACS}
U.~C. Bureau.
\newblock {AMERICAN COMMUNITY SURVEY 2012-2016 ACS 5-YEAR PUMS FILES}, 2018.

\bibitem{chan2012improved}
J.~C. Chan and D.~P. Kroese.
\newblock Improved cross-entropy method for estimation.
\newblock {\em Statistics and computing}, 22(5):1031--1040, 2012.

\bibitem{Chen2018}
W.~Y. Chen, L.~Mackey, J.~Gorham, F.-X. Briol, and C.~J. Oates.
\newblock {Stein points}.
\newblock In {\em Proceedings of the International Conference on Machine
  Learning, PMLR 80}, pages 843--852, 2018.

\bibitem{Chen2010}
Y.~Chen, M.~Welling, and A.~Smola.
\newblock {Super-samples from kernel herding}.
\newblock In {\em Proceedings of the Conference on Uncertainty in Artificial
  Intelligence}, 2012.

\bibitem{chipman2012sequential}
H.~Chipman, P.~Ranjan, and W.~Wang.
\newblock Sequential design for computer experiments with a flexible bayesian
  additive model.
\newblock {\em Canadian Journal of Statistics}, 40(4):663--678, 2012.

\bibitem{CART}
H.~A. Chipman, E.~I. George, and R.~E. McCulloch.
\newblock Bayesian {CART} model search.
\newblock {\em Journal of the American Statistical Association},
  93(443):935--948, 1998.

\bibitem{BART}
H.~A. Chipman, E.~I. George, and R.~E. McCulloch.
\newblock {BART}: {B}ayesian additive regression trees.
\newblock {\em Annals of Applied Statistics}, 4(1):266--298, 2010.

\bibitem{Chipman2012}
H.~A. Chipman, P.~Ranjan, and W.~Wang.
\newblock {Sequential design for computer experiments with a flexible Bayesian
  additive model}.
\newblock {\em Canadian Journal of Statistics}, 40(4):663--678, 2012.

\bibitem{Cockayne2017BPNM}
J.~Cockayne, C.~Oates, T.~Sullivan, and M.~Girolami.
\newblock {Bayesian probabilistic numerical methods}.
\newblock {\em SIAM Review}, 61(4):756--789, 2019.

\bibitem{Cornford1999}
D.~Cornford, I.~T. Nabney, and C.~K.~I. Williams.
\newblock {Adding constrained discontinuities to Gaussian process models of
  wind fields}.
\newblock {\em Advances in Neural Information Processing Systems}, pages
  861--867, 1999.

\bibitem{Davis2007}
P.~J. Davis and P.~Rabinowitz.
\newblock {\em {Methods of numerical integration}}.
\newblock Courier Corporation, 2007.

\bibitem{doi:10.1146/annurev-statistics-010814-020133}
A.~C. Davison and R.~Huser.
\newblock Statistics of extremes.
\newblock {\em Annual Review of Statistics and Its Application}, 2(1):203--235,
  2015.

\bibitem{Moral2006}
P.~Del~Moral, A.~Doucet, and A.~Jasra.
\newblock {Sequential Monte Carlo samplers}.
\newblock {\em Journal of the Royal Statistical Society Series B: Statisitical
  Methodology}, 68(3):411--436, 2006.

\bibitem{Denison1998}
D.~G.~T. Denison, B.~K. Mallick, and A.~F.~M. Smith.
\newblock {A Bayesian CART algorithm}.
\newblock {\em Biometrika}, 85(2):363--377, 1998.

\bibitem{Diaconis1988}
P.~Diaconis.
\newblock {Bayesian numerical analysis}.
\newblock {\em Statistical Decision Theory and Related Topics IV}, 1:163--175,
  1988.

\bibitem{Dick2013}
J.~Dick, F.~Y. Kuo, and I.~H. Sloan.
\newblock {High-dimensional integration: The quasi-Monte Carlo way}.
\newblock {\em Acta Numerica}, 22:133--288, 2013.

\bibitem{Ding2012}
J.~Ding, A.~Bashashati, A.~Roth, A.~Oloumi, K.~Tse, T.~Zeng, G.~Haffari,
  M.~Hirst, M.~A. Marra, A.~Condon, S.~Aparicio, and S.~P. Shah.
\newblock {Feature-based classifiers for somatic mutation detection in
  tumour-normal paired sequencing data}.
\newblock {\em Bioinformatics}, 28(2):167--175, 2012.

\bibitem{dorie2018dbarts}
V.~Dorie, H.~Chipman, and R.~McCulloch.
\newblock dbarts: Discrete {B}ayesian additive regression trees sampler, 2018.

\bibitem{dorie2019automated}
V.~Dorie, J.~Hill, U.~Shalit, M.~Scott, D.~Cervone, et~al.
\newblock Automated versus do-it-yourself methods for causal inference: Lessons
  learned from a data analysis competition.
\newblock {\em Statistical Science}, 34(1):43--68, 2019.

\bibitem{Durrett:2010:PTE:1869916}
R.~Durrett.
\newblock {\em Probability: theory and examples}.
\newblock Cambridge University Press, New York, NY, USA, 4th edition, 2010.

\bibitem{felip2019tree}
J.~Felip, N.~Ahuja, and O.~Tickoo.
\newblock Tree pyramidal adaptive importance sampling.
\newblock {\em arXiv preprint arXiv:1912.08434}, 2019.

\bibitem{Fisher2020}
M.~A. Fisher, C.~J. Oates, C.~Powell, and A.~Teckentrup.
\newblock {A locally adaptive Bayesian cubature method}.
\newblock In {\em Twenty Third International Conference on Artificial
  Intelligence and Statistic}, 2020.

\bibitem{foster2020model}
T.~Foster, C.~L. Lei, M.~Robinson, D.~Gavaghan, and B.~Lambert.
\newblock Model evidence with fast tree based quadrature.
\newblock {\em arXiv preprint arXiv:2005.11300}, 2020.

\bibitem{MARS}
J.~H. Friedman.
\newblock Multivariate adaptive regression splines.
\newblock {\em The Annals of Statistics}, 19(1):1--67, 1991.

\bibitem{fuglstad2015does}
G.-A. Fuglstad, D.~Simpson, F.~Lindgren, and H.~Rue.
\newblock Does non-stationary spatial data always require non-stationary random
  fields?
\newblock {\em Spatial Statistics}, 14:505--531, 2015.

\bibitem{gardner2018gpytorch}
J.~Gardner, G.~Pleiss, K.~Q. Weinberger, D.~Bindel, and A.~G. Wilson.
\newblock Gpytorch: Blackbox matrix-matrix {G}aussian process inference with
  {GPU} acceleration.
\newblock In {\em Advances in Neural Information Processing Systems}, pages
  7576--7586, 2018.

\bibitem{10.5555/3042573.3042683}
R.~Garnett, Y.~Krishnamurthy, X.~Xiong, J.~Schneider, and R.~Mann.
\newblock Bayesian optimal active search and surveying.
\newblock In {\em Proceedings of the 29th International Coference on
  International Conference on Machine Learning}, pages 843--850, 2012.

\bibitem{gelman2006data}
A.~Gelman and J.~Hill.
\newblock {\em Data analysis using regression and multilevel/hierarchical
  models}.
\newblock Cambridge university press, 2006.

\bibitem{Genz}
A.~Genz.
\newblock Testing multidimensional integration routines.
\newblock In {\em Proceedings of the International Conference on Tools, Methods
  and Languages for Scientific and Engineering Computation}, pages 81--94,
  1984.

\bibitem{george2019}
E.~George, P.~Laud, B.~Logan, R.~McCulloch, and R.~Sparapani.
\newblock {\em Fully nonparametric Bayesian additive regression trees}, pages
  89--110.
\newblock 10 2019.

\bibitem{Gessner2019}
A.~Gessner, J.~Gonzalez, and M.~Mahsereci.
\newblock {Active multi-information source Bayesian quadrature}.
\newblock In {\em Uncertainty in Artificial Intelligence}, 2019.

\bibitem{Ghosal2017}
S.~Ghosal and A.~van~der Vaart.
\newblock {\em {Fundamentals of nonparametric Bayesian inference}}.
\newblock Cambridge University Press, 2017.

\bibitem{Gunter2014}
T.~Gunter, R.~Garnett, M.~Osborne, P.~Hennig, and S.~Roberts.
\newblock {Sampling for inference in probabilistic models with fast Bayesian
  quadrature}.
\newblock In {\em Advances in Neural Information Processing Systems}, pages
  2789--2797, 2014.

\bibitem{hahn2020}
P.~R. Hahn, J.~S. Murray, and C.~M. Carvalho.
\newblock Bayesian regression tree models for causal inference: Regularization,
  confounding, and heterogeneous effects (with discussion).
\newblock {\em Bayesian Anal.}, 15(3):965--1056, 09 2020.

\bibitem{hanson2011polya}
T.~E. Hanson, J.~V. Monteiro, and A.~Jara.
\newblock The polya tree sampler: Toward efficient and automatic independent
  metropolis--hastings proposals.
\newblock {\em Journal of Computational and Graphical Statistics},
  20(1):41--62, 2011.

\bibitem{MH}
W.~K. Hastings, editor.
\newblock {\em Classification, subtype discovery, numerical recipes: The art of
  scientific computing (3rd Edition)}.
\newblock Oxford University Press on behalf of Biometrika Trust, New York,
  2007.

\bibitem{he2019xbart}
J.~He, S.~Yalov, and P.~R. Hahn.
\newblock {XBART: Accelerated Bayesian additive regression trees}.
\newblock In {\em The 22nd International Conference on Artificial Intelligence
  and Statistics}, pages 1130--1138, 2019.

\bibitem{Hennig2015}
P.~Hennig, M.~A. Osborne, and M.~Girolami.
\newblock {Probabilistic numerics and uncertainty in computations}.
\newblock {\em Journal of the Royal Society A}, 471(2179):20150142, 2015.

\bibitem{Hill2020}
J.~Hill, A.~Linero, and J.~Murray.
\newblock {Bayesian additive regression trees: a review and look forward}.
\newblock {\em Annual Reviews in Statistics and Its Applications}, 7(6):1--28,
  2020.

\bibitem{Hill2011}
J.~L. Hill.
\newblock {Bayesian nonparametric modeling for causal inference}.
\newblock {\em Journal of Computational and Graphical Statistics},
  20(1):217--240, 2011.

\bibitem{Jagadeeswaran2018}
R.~Jagadeeswaran and F.~J. Hickernell.
\newblock {Fast automatic Bayesian cubature using lattice sampling}.
\newblock {\em Statistics and Computing}, 29(6):1215--1229, 2019.

\bibitem{jeong_art_2020}
S.~Jeong and V.~Rockova.
\newblock The art of bart: On flexibility of bayesian forests.
\newblock {\em arXiv preprint arXiv:2008.06620}, 2020.

\bibitem{Jiang2019}
S.~Jiang, H.~Chai, J.~Gonzalez, and R.~Garnett.
\newblock {BINOCULARS for efficient, nonmyopic sequential experimental design}.
\newblock {\em arXiv:1909.04568}, 2019.

\bibitem{Kanagawa2019}
M.~Kanagawa and P.~Hennig.
\newblock {Convergence guarantees for adaptive Bayesian quadrature methods}.
\newblock In {\em Advances in Neural Information Processing Systems}, pages
  6234--6245, 2019.

\bibitem{Kanagawa2017}
M.~Kanagawa, B.~K. Sriperumbudur, and K.~Fukumizu.
\newblock {Convergence analysis of deterministic kernel-based quadrature rules
  in misspecified settings}.
\newblock {\em Foundations of Computational Mathematics}, pages 1--40, 2019.

\bibitem{Karvonen2017symmetric}
T.~Karvonen and S.~S{\"{a}}rkk{\"{a}}.
\newblock Fully symmetric kernel quadrature.
\newblock {\em SIAM Journal on Scientific Computing}, 40(2):A697--A720, 2018.

\bibitem{Karvonen2019}
T.~Karvonen, S.~S{\"{a}}rkk{\"{a}}, and C.~J. Oates.
\newblock {Symmetry exploits for Bayesian cubature methods}.
\newblock {\em Statistics and Computing}, 29:1231--1248, 2019.

\bibitem{Kroese2014WhyTM}
D.~P. Kroese, T.~J. Brereton, T.~Taimre, and Z.~I. Botev.
\newblock Why the monte carlo method is so important today.
\newblock {\em Wiley Interdisciplinary Reviews: Computational Statistics},
  6:386--392, 2014.

\bibitem{linero2018dirichlet}
A.~R. L.
\newblock Bayesian regression trees for high-dimensional prediction and
  variable selection.
\newblock {\em Journal of the American Statistical Association},
  113(522):626--636, 2018.

\bibitem{Lakshminarayanan2016}
B.~Lakshminarayanan, D.~M. Roy, and Y.~W. Teh.
\newblock {Mondrian forests for large-scale regression when uncertainty
  matters}.
\newblock {\em Proceedings of the 19th International Conference on Artificial
  Intelligence and Statistics}, 51:1478--1487, 2016.

\bibitem{Larkin1972}
F.~M. Larkin.
\newblock {Gaussian measure in Hilbert space and applications in numerical
  analysis}.
\newblock {\em Rocky Mountain Journal of Mathematics}, 2(3):379--422, 1972.

\bibitem{PeterLepage1978}
G.~P. Lepage.
\newblock {A new algorithm for adaptive multidimensional integration}.
\newblock {\em Journal of Computational Physics}, 27(2), 1978.

\bibitem{AMI}
G.~P. Lepage.
\newblock A new algorithm for adaptive multidimensional integration.
\newblock {\em Journal of Computational Physics}, 27(2):192--203, 1978.

\bibitem{lepage2020adaptive}
G.~P. Lepage.
\newblock Adaptive multidimensional integration: Vegas enhanced.
\newblock {\em arXiv preprint arXiv:2009.05112}, 2020.

\bibitem{linero2018bayesian}
A.~R. Linero and Y.~Yang.
\newblock Bayesian regression tree ensembles that adapt to smoothness and
  sparsity.
\newblock {\em Journal of the Royal Statistical Society: Series B (Statistical
  Methodology)}, 80(5):1087--1110, 2018.

\bibitem{Liu2016SVGD}
Q.~Liu and D.~Wang.
\newblock {Stein variational gradient descent: A general purpose Bayesian
  inference algorithm}.
\newblock In {\em Advances In Neural Information Processing Systems}, pages
  2370--2378, 2016.

\bibitem{llorente2020adaptive}
F.~Llorente, L.~Martino, V.~Elvira, D.~Delgado, and J.~Lopez-Santiago.
\newblock Adaptive quadrature schemes for bayesian inference via active
  learning.
\newblock {\em arXiv preprint arXiv:2006.00535}, 2020.

\bibitem{Meyn1993}
S.~Meyn and R.~L. Tweedie.
\newblock {\em Markov Chains and Stochastic Stability}.
\newblock Springer, 1993.

\bibitem{Mohammadi2019}
H.~Mohammadi, P.~Challenor, M.~Goodfellow, and D.~Williamson.
\newblock Emulating computer models with step-discontinuous outputs using
  {G}aussian processes.
\newblock {\em arXiv:1903.02071}, 2019.

\bibitem{moon1996expectation}
T.~K. Moon.
\newblock The expectation-maximization algorithm.
\newblock {\em IEEE Signal processing magazine}, 13(6):47--60, 1996.

\bibitem{network}
A.~S. Naeem, M.
\newblock Kegg metabolic reaction network (undirected) data set.

\bibitem{oates2019bayesian}
C.~J. Oates, J.~Cockayne, R.~G. Aykroyd, and M.~Girolami.
\newblock Bayesian probabilistic numerical methods in time-dependent state
  estimation for industrial hydrocyclone equipment.
\newblock {\em Journal of the American Statistical Association},
  114(528):1518--1531, 2019.

\bibitem{Oates2016CF2}
C.~J. Oates, J.~Cockayne, F.-X. Briol, and M.~Girolami.
\newblock {Convergence rates for a class of estimators based on Stein's
  identity}.
\newblock {\em Bernoulli}, 25(2):1141--1159, 2019.

\bibitem{Oates2017}
C.~J. Oates, M.~Girolami, and N.~Chopin.
\newblock {Control functionals for Monte Carlo integration}.
\newblock {\em Journal of the Royal Statistical Society B: Statistical
  Methodology}, 79(3):695--718, 2017.

\bibitem{Oates2017heart}
C.~J. Oates, S.~Niederer, A.~Lee, F.-X. Briol, and M.~Girolami.
\newblock {Probabilistic models for integration error in the assessment of
  functional cardiac models}.
\newblock In {\em Advances in Neural Information Processing}, pages 110--118,
  2017.

\bibitem{Oettershagen2017}
J.~Oettershagen.
\newblock {\em Construction of optimal cubature algorithms with applications to
  econometrics and uncertainty quantification}.
\newblock PhD thesis, Rheinischen Friedrich-Wilhelms-Universit{\"{a}}t Bonn,
  2017.

\bibitem{ohagan1991bayes}
A.~O'Hagan.
\newblock Bayes--hermite quadrature.
\newblock {\em Journal of Statistical Planning and Inference}, 29(3):245--260,
  1991.

\bibitem{OHagan1991}
A.~O'Hagan.
\newblock {Bayes–Hermite quadrature}.
\newblock {\em Journal of Statistical Planning and Inference}, 29(3):245--260,
  1991.

\bibitem{OHagan1992}
A.~O'Hagan.
\newblock {Some Bayesian numerical analysis}.
\newblock {\em Bayesian Statistics}, 4:345--363, 1992.

\bibitem{o2018scalable}
M.~O'Kelly, A.~Sinha, H.~Namkoong, R.~Tedrake, and J.~C. Duchi.
\newblock Scalable end-to-end autonomous vehicle testing via rare-event
  simulation.
\newblock In {\em Advances in Neural Information Processing Systems}, pages
  9827--9838, 2018.

\bibitem{GPSQ}
M.~A. Osborne.
\newblock {\em Bayesian Gaussian processes for sequential prediction,
  optimisation and quadrature}.
\newblock PhD thesis, Oxford University, 2010.

\bibitem{Paulin2015}
D.~Paulin.
\newblock {Concentration inequalities for Markov chains by Marton couplings and
  spectral methods}.
\newblock {\em Electronic Journal of Probability}, 20(79):1--32, 2015.

\bibitem{Polson2018}
N.~Polson and V.~Rockova.
\newblock {Posterior concentration for sparse deep learning}.
\newblock In {\em Advances in Neural Information Processing Systems}, pages
  930--941, 2018.

\bibitem{Pratola2016}
M.~T. Pratola.
\newblock {Efficient Metropolis-Hastings proposal mechanisms for Bayesian
  regression tree models}.
\newblock {\em Bayesian Analysis}, 11(3):885--911, 2016.

\bibitem{Pratola2014}
M.~T. Pratola, H.~A. Chipman, J.~R. Gattiker, D.~M. Higdon, R.~McCulloch, and
  W.~N. Rust.
\newblock {Parallel Bayesian additive regression trees}.
\newblock {\em Journal of Computational and Graphical Statistics},
  23(3):830--852, 2014.

\bibitem{pratola2017}
M.~T. Pratola, H.~A. Chipman, E.~George, and R.~McCulloch.
\newblock Heteroscedastic {BART} using multiplicative regression trees.
\newblock {\em Journal of Computational and Graphical Statistics}, 09 2017.

\bibitem{BQ}
S.~A. V. W. T. F. B.~P. Press, W.H.;~Teukolsky, editor.
\newblock {\em Classification, subtype discovery, numerical recipes: The art of
  scientific computing (3rd Edition)}.
\newblock Cambridge University Press, New York, 2007.

\bibitem{Press:2007:NRE:1403886}
W.~H. Press, S.~A. Teukolsky, W.~T. Vetterling, and B.~P. Flannery.
\newblock {\em Numerical recipes 3rd edition: The art of scientific computing.}
\newblock Cambridge University Press, New York, NY, USA, 3 edition, 2007.

\bibitem{Quann2018}
M.~Quann, L.~Ojeda, W.~Smith, D.~Rizzo, M.~Castanier, and K.~Barton.
\newblock {Ground robot terrain mapping and energy prediction in environments
  with 3-D topography}.
\newblock {\em Proceedings of the American Control Conference}, pages
  3532--3537, 2018.

\bibitem{R}
{R Core Team}.
\newblock {\em R: A language and environment for statistical computing}.
\newblock R Foundation for Statistical Computing, Vienna, Austria, 2018.

\bibitem{Rasmussen2003}
C.~Rasmussen and Z.~Ghahramani.
\newblock {Bayesian Monte Carlo}.
\newblock In {\em Advances in Neural Information Processing Systems}, pages
  489--496, 2002.

\bibitem{Rasmussen2006}
C.~Rasmussen and C.~Williams.
\newblock {\em {Gaussian Processes for Machine Learning}}.
\newblock MIT Press, 2006.

\bibitem{Rasmussen:2002:BMC:2968618.2968681}
C.~E. Rasmussen and Z.~Ghahramani.
\newblock {Bayesian Monte Carlo}.
\newblock In {\em Advances in Neural Information Processing Systems}, pages
  505--512, 2002.

\bibitem{Rasmussen:2005:GPM:1162254}
C.~E. Rasmussen and C.~K.~I. Williams.
\newblock {\em Gaussian processes for machine learning (Adaptive computation
  and machine learning)}, volume~2.
\newblock The MIT Press, 2006.

\bibitem{Robert2004}
C.~Robert and G.~Casella.
\newblock {\em {Monte Carlo statistical methods}}.
\newblock Springer, 2004.

\bibitem{Roberts2004}
G.~O. Roberts and J.~S. Rosenthal.
\newblock {General state space Markov chains and MCMC algorithms}.
\newblock {\em Probability Surveys}, 1(1):20--71, 2004.

\bibitem{Rockova2019BvM}
V.~Rockova.
\newblock {On semi-parametric Bernstein-von Mises theorems for BART}.
\newblock {\em arXiv:1905.03735}, 2019.

\bibitem{rockova2019theory}
V.~Rockova and E.~Saha.
\newblock {On theory for BART}.
\newblock In {\em Proceedings of Machine Learning Research, PMLR 89}, pages
  2839--2848, 2019.

\bibitem{rockova2017posterior}
V.~Rockova and S.~van~der Pas.
\newblock {Posterior concentration for Bayesian regression trees and forests}.
\newblock {\em The Annals of Statistics}, 2017.

\bibitem{rubino2009rare}
G.~Rubino and B.~Tuffin.
\newblock {\em Rare event simulation using Monte Carlo methods}.
\newblock John Wiley \& Sons, 2009.

\bibitem{rubinstein2004cross}
R.~Y. Rubinstein and D.~P. Kroese.
\newblock The cross-entropy method: A unified approach to monte carlo
  simulation, randomized optimization and machine learning.
\newblock {\em Information Science \& Statistics, Springer Verlag, NY}, 2004.

\bibitem{Rue2016}
H.~Rue, A.~Riebler, S.~H. S{\o}rbye, J.~B. Illian, D.~P. Simpson, and F.~K.
  Lindgren.
\newblock {Bayesian computing with INLA: A review}.
\newblock {\em Annual Reviews of Statistics and Its Applications}, 4:395--421,
  2016.

\bibitem{Santner2018}
T.~J. Santner, B.~J. Williams, and W.~I. Notz.
\newblock {\em {The design and analysis of computer experiments}}.
\newblock Springer, 2nd edition, 2018.

\bibitem{rescallingGenz}
R.~Sch{\"u}rer.
\newblock {Parallel High-dimensional integration: quasi-Monte Carlo versus
  adaptive cubature rules}.
\newblock In V.~N. Alexandrov, J.~J. Dongarra, B.~A. Juliano, R.~S. Renner, and
  C.~J.~K. Tan, editors, {\em Computational Science --- ICCS 2001}, pages
  1262--1271, Berlin, Heidelberg, 2001. Springer Berlin Heidelberg.

\bibitem{7352306}
B.~Shahriari, K.~Swersky, Z.~Wang, R.~P. Adams, and N.~de~Freitas.
\newblock Taking the human out of the loop: A review of bayesian optimization.
\newblock {\em Proceedings of the IEEE}, 104(1):148--175, 2016.

\bibitem{Suldin1959}
A.~B. Suldin.
\newblock {Wiener measure and its applications to approximation methods. I}.
\newblock {\em I. Izvestiya Vysshikh Uchebnykh Zavedenii. Matematika}, 1959(6
  (13)):145--158, 1959.

\bibitem{Taddy2011}
M.~A. Taddy, R.~B. Gramacy, and N.~G. Polson.
\newblock Dynamic trees for learning and design.
\newblock {\em Journal of the American Statistical Association},
  106(493):109--123, 2011.

\bibitem{Teckentrup2019}
A.~L. Teckentrup.
\newblock {Convergence of Gaussian process regression with estimated
  hyper-parameters and applications in Bayesian inverse problems}.
\newblock {\em arXiv:1909.00232}, 2019.

\bibitem{us_census_bureau_american_2018}
{U.S. Census Bureau}.
\newblock American {Community} {Survey}, 2012-2016 {ACS} 5-{Year} {PUMS}
  {Files}.
\newblock Technical report, U.S. Department of Commerce, Janurary 2018.

\bibitem{Vaart2011}
A.~{van Der Vaart} and H.~van Zanten.
\newblock {Information rates of nonparametric Gaussian process methods}.
\newblock {\em Journal of Machine Learning Research}, 12(Jun):2095--2119, 2011.

\bibitem{10.2307/25464673}
A.~W. van~der Vaart and J.~H. van Zanten.
\newblock Rates of contraction of posterior distributions based on gaussian
  process priors.
\newblock {\em The Annals of Statistics}, 36(3):1435--1463, 2008.

\bibitem{Wang2020}
Y.~Wang and V.~Ro{\v{c}}kov{\'{a}}.
\newblock {Uncertainty quantification for sparse deep learning}.
\newblock {\em Artificial Intelligence and Statistics}, 2020.

\bibitem{wilson2018maximizing}
J.~T. Wilson, F.~Hutter, and M.~P. Deisenroth.
\newblock Maximizing acquisition functions for bayesian optimization, 2018.

\bibitem{Wynne2020}
G.~Wynne, F.-X. Briol, and M.~Girolami.
\newblock {Convergence guarantees for Gaussian process means with misspecified
  likelihoods and smoothness}.
\newblock {\em arXiv:2001.10818}, 2020.

\bibitem{Xi2018MultiOutput}
X.~Xi, F.-X. Briol, and M.~Girolami.
\newblock {Bayesian quadrature for multiple related integrals}.
\newblock In {\em International Conference on Machine Learning, PMLR 80}, pages
  5369--5378, 2018.

\bibitem{yeager2019national}
D.~S. Yeager, P.~Hanselman, G.~M. Walton, J.~S. Murray, R.~Crosnoe, C.~Muller,
  E.~Tipton, B.~Schneider, C.~S. Hulleman, C.~P. Hinojosa, et~al.
\newblock A national experiment reveals where a growth mindset improves
  achievement.
\newblock {\em Nature}, 573(7774):364--369, 2019.

\bibitem{zhu_bayesian_2020}
H.~Zhu, X.~Liu, R.~Kang, Z.~Shen, S.~Flaxman, and F.-X. Briol.
\newblock Bayesian {Probabilistic} {Numerical} {Integration} with
  {Tree}-{Based} {Models}.
\newblock {\em arXiv preprint arXiv:2006.05371}, 2020.

\end{thebibliography}


\onecolumn

\appendix


\begin{center}

\vspace{0.3cm}
\LARGE \textbf{Appendices} 
\vspace{0.3cm}
\end{center}

This supplementary material is separated into four sections. First, in Appendix \ref{appendix:BART}, we provided a detailed description of the BART prior considered in the paper. Secondly, in Appendix \ref{appendix:posterior} we give a concise description of the posterior sampling procedure. Then, in Appendix \ref{appendix:proofs}, we provide proofs for the theoretical result in the main text. Finally, in Appendix \ref{appendix:experiments}, we provide additional details on the experimental setting as well as additional numerical results.

\section{The Bayesian Additive Regression Trees Prior} \label{appendix:BART}
Following our definition in Section \ref{BART}, for a fixed number of regression trees $T$, a \textit{T-additive regression tree} $g_{\mathcal{E}, \mathcal{B}}$ is essentially determined by $(\mathcal{T}_1, \beta_1), \ldots, (\mathcal{T}_T, \beta_T)$, where for each tree $t$, $\mathcal{T}_t$ is a partition of $\mathcal{X}$ into $K_t$ subsets and $\beta_t$ is a vector of leaf values. Additionally, $\sigma$ is the standard deviance of the Gaussian observational noise. This section summarizes the discussion in \cite{BART} on how the prior $\P((\mathcal{T}_1, \beta_1), \ldots, (\mathcal{T}_T, \beta_T), \sigma)$ is chosen. An appropriate prior can effectively prevent the individual trees from being overly influential, thus regularizing the fit as an ensemble model. We refer the reader to \cite{BART} for more details. Details about our specific choice of hyper-parameters for the priors can be found in Section~\ref{appendix:genz}.

We will assume independence amongst tree components $(\mathcal{T}_t, \beta_t)$ and $\sigma$, and amongst each tree's terminal node parameters $\beta_t$. The distribution of the sum-of-trees model can hence be simplified as:
\begin{talign*}
p((\mathcal{T}_1, \beta_1), \ldots, (\mathcal{T}_T, \beta_T), \sigma) & = p(\sigma) \prod_{t} p(\beta_t|\mathcal{T}_t) p(\mathcal{T}_t)\\
p(\beta_t|\mathcal{T}_t) & = \prod_{k} p(\beta_{t, k}|\mathcal{T}_t),
\end{talign*}
where $\beta_{t, k}$ is the $k$-th terminal node of the $t$-th tree. We further assume an identical form for each of the component $p(\mathcal{T}_t)$ and for $p(\beta_t | \mathcal{T}_t)$. The prior distributions to be specified are therefore $p(\mathcal{T}_t)$, $p(\beta_t | \mathcal{T}_t)$ and $p(\sigma)$.

\paragraph{The  \texorpdfstring{$\mathcal{T}_t$}{T} Prior}

The trees in BART are $k$-d trees and have axis-aligned splits. For the prior on $\mathcal{T}_t$, we will use what is commonly known as the ``uniform'' prior in the literature. This is usually specified in the form of a generative model which consists of three components. Firstly, we specify the probability that a given node of depth $l \in \mathbb{N} \cup \{0\}$ is terminal. This takes the form $p_\text{split} = \alpha (1 + l)^{-\beta}$, where $\alpha \in (0, 1)$ and $\beta \in [0, \infty)$. Secondly, a uniform distribution on $\{1,\ldots,d\}$ is used to decide which of the available variables $x_1,\ldots,x_d$ to split on at each interior node. Lastly, a uniform distribution on the set of possible values of that variable is used for the splitting rule assignment. See \cite{dorie2018dbarts, CART} for more details. Other priors are also possible, such as the Galton-Watson prior \cite{rockova2019theory}.

\paragraph{The  \texorpdfstring{$\beta_{t} | \mathcal{T}_t$}{beta conditional T} Prior}
\label{Parameter Prior 2}
For the distribution of leaf values given a tree, we use a Gaussian distribution: 
\begin{talign*}
p(\beta_{t}|\mathcal{T}_t) = \prod_{k=1}^{K_t}
N\left(\beta_{t,k}; 0, \sigma_\beta^2\right).
\end{talign*}
Let $y_{\text{min}} = \min_{i} y^n_i$ and $y_{\text{max}} = \max_{i} y^n_i$. In \cite{BART}, the authors suggested to rescale $y$ to have zero mean and to ensure that $y_{\text{min}} = -0.5$ and $y_{\text{max}} = 0.5$, then taking $\sigma_{\beta} = 0.25 /\sqrt{T}$. This ensures that the prior on $g(x)$ is within the interval $(y_\text{min}, y_\text{max})$ with high probability. 

\paragraph{The  \texorpdfstring{$\sigma$}{sigma} Prior}

For the prior on $\sigma$, we use the inverse chi-square distribution $\sigma^{2} \sim \nu \lambda / \chi_{\nu}^2$. This is also a conjugate prior which, again, reduces the required computational effort in the MCMC procedure, as elaborated in \cite{BART}. To find the appropriate hyper-parameters, we introduce $q \in (0, 1)$, take $\hat{\sigma}$ of $\sigma$ as the sample standard deviation of $y^n$ and calibrate for $\nu$ and $\lambda$ such that $\mathbb{P}(\sigma < \hat{\sigma}) = q$. The authors of \cite{BART} recommended $(\nu,q)=(3,0.90)$ as it tends to avoid extremes and we hence follow these recommendations.

\paragraph{Number of Trees  \texorpdfstring{$T$}{T}}
Finally, the number of regression trees $T$ can either be chosen to be the default value $T=200$, or through cross-validation. \cite{BART} further pointed out that, in general, the model performs reasonably well on prediction tasks so long as the value of $T$ is not too small.

\section{Bayesian Additive Regression Trees Posterior Sampling Procedure}
\label{appendix:posterior}
In this appendix, we now describe the Bayesian backfitting MCMC algorithm first introduced in \cite{BART}, and which is used throughout our experiments. We will make use of the notation: $\mathcal{T}_{-j}=\mathcal{E}\setminus \{\mathcal{T}_j\}$ and similarly $\beta_{-j}=\mathcal{B}\setminus \{\beta_j\}$ for $j=1,\ldots,T$. Our target posterior is the distribution $\mathcal{E},\mathcal{B}, \sigma| y^n$. To sample from this posterior, we will make use of a Gibbs sampler, which, at each iteration, draws $(\mathcal{T}_j,\beta_j) | ( \mathcal{T}_{-j},\beta_{-j},\sigma,y^n )$ sequentially for $t = 1, \ldots, T$, then draws $\sigma$ from  $\sigma|(\mathcal{E}, \mathcal{B},y^n)$.
\paragraph{Sampling individual trees} We note that drawing from $(\mathcal{T}_j,\beta_j) | (\mathcal{T}_{-j},\beta_{-j},\sigma,y^n)$ is equivalent to drawing from
\begin{talign*}
(\mathcal{T}_j,\beta_j) | R_j,\sigma, \qquad
\text{where } \qquad (R_j)_k= y_k - \sum_{t\neq j} g_{\mathcal{T}_t, {\beta}_t}(x_k),
\end{talign*}
for $k=1,\ldots,n$, which is the partial residual obtained without the $j$-th tree. Drawing from this distribution is the same as drawing from the posterior of the residuals regression model with the $j$-th tree $R_j = g_{\mathcal{T}_j, \beta_j}(x) + \epsilon$, where $\epsilon\sim\mathcal{N}(0,\sigma^2)$. Since the prior distribution $p(\beta_j)$ and the likelihood $p(R_j|\beta_j,\mathcal{T}_j,\sigma)$ are Gaussian, the posterior distribution attains a closed form up to a normalising constant 
\begin{talign*}
p(\mathcal{T}_j|R_j,\sigma) \propto p(\mathcal{T}_j)\int_{\mathbb{R}^{K_j}} p(R_j|\beta_j,\mathcal{T}_j,\sigma)p(\beta_j|\mathcal{T}_j,\sigma) d\beta_j,
\end{talign*}
where $d\beta_j$ is the Lebesgue measure on $\mathbb{R}^{K_j}$ and $K_j$ is the number of leaf nodes for tree $j$. Therefore we can now sample $\mathcal{T}_j$ and $\beta_j$ via the following two-step procedure: first sample from $\mathcal{T}_j |R_j,\sigma$, then sample from $\beta_j | \mathcal{T}_j,R_j,\sigma$.

The draw of $\mathcal{T}_j | R_j,\sigma$ is done via a Metropolis-Hastings algorithm with the following proposal. Given the current tree, grow a terminal node with probability $0.25$, prune a pair of terminal nodes with probability $0.25$, change a non-terminal node's split rule with probability $0.40$, and finally swap a split rule between parent and child with probability $0.10$.

\paragraph{Sampling the standard deviation of the observational noise} To sample from $\sigma|(\mathcal{E}, \mathcal{B},y^n)$, we simply draw from the inverse chi-squared distribution defined in Appendix \ref{appendix:BART}.

\section{Proofs of Theoretical Results} \label{appendix:proofs}

In this appendix, we present the proofs of our theoretical results. Section \ref{sec:proof_MCMCsamples} provides a proof of Proposition \ref{eq:BARTBQ_mean&var}, Section \ref{sec:convergence_proof} provides a proof of Theorem \ref{thm:convergence_rate}, and Section \ref{sec:MCMC_proof} provides a proof of Proposition \ref{prop:MCMC_convergence}.

\subsection{Proof of Proposition \ref{eq:BARTBQ_mean&var}}\label{sec:proof_MCMCsamples}

\begin{proof}
We begin by deriving the integral of some arbitrary $T$-additive tree $g_{\mathcal{E},\mathcal{B}}:\mathcal{X} \times \Omega \rightarrow \mathbb{R}$ against the probability measure $\Pi$ on $\mathcal{X}$. Fix $\omega \in \Omega$. Then, using linearity of integration, we get:
\begin{talign*}
    \Pi[g_{\mathcal{E},\mathcal{B}}(\cdot,\omega)] & = \int_{\mathcal{X}} g_{\mathcal{E},\mathcal{B}}(x,\omega) d\Pi(x)  = \int_{\mathcal{X}} \sum_{t=1}^T\sum_{k=1}^{K} \beta_{t,k}  \mathbbm{1}_{\chi_{t, k}}(x)d\Pi(x)\\
    &
    = \sum_{t=1}^T\sum_{k=1}^{K} \beta_{t,k} \int_{\mathcal{X}} \mathbbm{1}_{\chi_{t, k}}(x)d\Pi(x) 
    = \sum_{t=1}^T\sum_{k=1}^{K} \beta_{t,k} \Pi(\chi_{t, k}).
\end{talign*}
We can use this expression to derive the posterior mean and variance for $\Pi[g]$ given the data $X,y$:
\begin{talign*}
    \mathbb{E}[\Pi[g_{\mathcal{E},\mathcal{B}}]|X,y] &  = \mathbb{E}\left[\sum_{t=1}^T\sum_{k=1}^{K} \beta_{t,k} \Pi(\chi_{t, k})\Big|X,y\right], \\
    \mathbb{V}[\Pi[g_{\mathcal{E},\mathcal{B}}]|X,y] &= \mathbb{E}\left[ \left(\Pi[g_{\mathcal{E},\mathcal{B}}\right] - \mathbb{E}\left[\Pi\left[g_{\mathcal{E},\mathcal{B}}]|X,y\right]\right)^2\Big| X,y \right].
\end{talign*}
Using a U-statistic estimate of these quantities based on posterior samples for the parameters leads to our desired result.
\end{proof}


\subsection{Proof of Theorem \ref{thm:convergence_rate}}
\label{sec:convergence_proof}

\begin{proof}
Fixing $\omega \in \Omega$ (i.e. fixing a realisation from the stochastic process) and starting with the triangle inequality, we can decouple the integration error into several terms depending on $g(\cdot,\omega)$:
\begin{talign}
    & | \Pi[f] - \Pi[g(\cdot,\omega)]| \nonumber \\
    & =
    \left|\Pi[f] - \Pi[g(\cdot,\omega)] + \frac{1}{n}\sum_{i=1}^n g(x_i,\omega) - \frac{1}{n}\sum_{i=1}^n g(x_i,\omega) + \frac{1}{n}\sum_{i=1}^n f(x_i) - \frac{1}{n}\sum_{i=1}^n f(x_i) \right|\nonumber \\
    & \leq
    \left|\Pi[f] - \frac{1}{n}\sum_{i=1}^n f(x_i)\right|
    + \left|\Pi[g(\cdot,\omega)]   - \frac{1}{n}\sum_{i=1}^n g(x_i,\omega) \right|\nonumber \\ & \qquad 
    + \left| \frac{1}{n}\sum_{i=1}^n g(x_i,\omega) - \frac{1}{n}\sum_{i=1}^n f(x_i) \right|.
    \label{eq:triangle_ineq_1}
\end{talign}
Note that by definition of the worst-case integration error in $\mathcal{H}$, we have that for any $h \in \mathcal{H}$:
\begin{talign}\label{eq:WCE}
    \left|\Pi[h] - \frac{1}{n}\sum_{i=1}^n h(x_i)\right| \leq \|h\|_{\mathcal{H}} \times \sup_{\|h\|_{\mathcal{H}}\leq 1} \left|\Pi[h] - \frac{1}{n}\sum_{i=1}^n h(x_i)\right|
\end{talign}
First, using Equation \ref{eq:WCE} in Equation \ref{eq:triangle_ineq_1}, then using assumption A2 gives us that whenever $n\geq N$, $\exists B>0$ such that:
\begin{talign}
    \left|\Pi[f] - \Pi[g(\cdot,\omega)] \right| 
    & \leq
    B\|f\|_{\mathcal{H}} \gamma_n
    + B\|g(\cdot,\omega)\|_{\mathcal{H}} \gamma_n
    + \left| \frac{1}{n}\sum_{i=1}^n g(x_i,\omega) - \frac{1}{n}\sum_{i=1}^n f(x_i) \right| \nonumber\\
     & =
    B (\|f\|_{\mathcal{H}} +\|g(\cdot,\omega)\|_{\mathcal{H}}) \gamma_n
    + \left| \frac{1}{n}\sum_{i=1}^n g(x_i,\omega) - \frac{1}{n}\sum_{i=1}^n f(x_i) \right|. \label{eq:pointset_bound}
\end{talign}
To tackle the third term, we can use the Cauchy-Schwartz inequality, which states that $\forall u,v \in \mathbb{R}^n$, we have $(\sum_{i=1}^n u_i v_i)^2 \leq (\sum_{i=1}^n u_i^2) (\sum_{i=1}^n v_i^2)$. Taking $u_i = g(x_i,\omega)-f(x_i)$ and $v_i=1$, we get:
\begin{talign*}
    \left( \sum_{i=1}^n g(x_i,\omega) -\sum_{i=1}^n f(x_i) \right)^2
    & \leq
    n \sum_{i=1}^n (g(x_i,\omega)-f(x_i))^2.
\end{talign*}
Multiplying both sides by $n^{-2}$ and taking square roots, we end up with:
\begin{talign}
    \left| \frac{1}{n} \sum_{i=1}^n g(x_i,\omega) - \frac{1}{n} \sum_{i=1}^n f(x_i)\right| 
    & \leq
    \left(\frac{1}{n}\sum_{i=1}^n (g(x_i,\omega) - f(x_i))^2\right)^{\frac{1}{2}}
    = 
    \|g(\cdot,\omega)-f\|_n. \label{eq:empiricalnorm_bound}
\end{talign}
Plugging in Equation \ref{eq:empiricalnorm_bound} into Equation \ref{eq:pointset_bound}, we get:
\begin{talign*}
    \left|\Pi[f] - \Pi[g(\cdot,\omega)] \right| 
    & \leq
    B (\|f\|_{\mathcal{H}} +\|g(\cdot,\omega)\|_{\mathcal{H}})\gamma_n
    + \|g(\cdot,\omega)-f\|_n \\
    & \leq
    B(\|f\|_{\mathcal{H}} +\|g(\cdot,\omega)\|_{\mathcal{H}}) \max(\varepsilon_n,\gamma_n)
    + \|g(\cdot,\omega)-f\|_n,
\end{talign*}
Using this inequality, we have that:
\begin{talign}
    \mathbb{P}_n&\left(| \Pi[f] -  \Pi[g(\cdot,\omega)]| > C_n \max(\varepsilon_n,\gamma_n) \right) \nonumber \\
    & \leq
    \mathbb{P}_n \left( B (\|f\|_{\mathcal{H}} +\|g(\cdot,\omega)\|_{\mathcal{H}}) \max(\varepsilon_n,\gamma_n)
    + \|g(\cdot,\omega)-f\|_n \geq C_n \max(\varepsilon_n,\gamma_n)\right) \nonumber \\
    & \leq
    \mathbb{P}_n\left( \|g-f\|_n \geq (C_n - B(\|f\|_{\mathcal{H}} +\|g(\cdot,\omega)\|_{\mathcal{H}})) \max(\varepsilon_n ,\gamma_n) \right) \nonumber \\
    & = \mathbb{P}_n\left( \|g-f\|_n \geq A_n \max(\varepsilon_n,\gamma_n) \right) \nonumber \\
    & \leq \mathbb{P}_n\left( \|g-f\|_n \geq A_n \varepsilon_n  \right),\label{eq:thm_final_bound}
\end{talign}
where we have taken $A_n=C_n - B(\|f\|_{\mathcal{H}} +\|g(\cdot,\omega)\|_{\mathcal{H}})$.
Clearly, since $C_n \rightarrow \infty$ as $n \rightarrow \infty$ and $B, \|f\|_{\mathcal{H}},\|g(\cdot,\omega)\|_{\mathcal{H}} <\infty$, we must have $A_n \rightarrow \infty$ as $n\rightarrow \infty$. Hence, combining Assumption A1 and the upper bound of Equation \ref{eq:thm_final_bound}, we have: 
\begin{talign*}
    \lim_{n\rightarrow \infty} \mathbb{P}_n[| \Pi[f] -  \Pi[g]| > C_n \max(\varepsilon_n,\gamma_n)] = 0
\end{talign*}
which concludes our proof.

\end{proof}


\subsection{Proof of Proposition \ref{prop:MCMC_convergence} and Discussion}
\label{sec:MCMC_proof}

\begin{proof}
Theorem 25 of \cite{Roberts2004} states that if a Markov chain with stationary distribution $\mathbb{P}$ is reversible and geometrically ergodic, then a central limit theorem holds for any $h\in L^2(\mathbb{P})$. Taking $h=\Pi[g]$, which is in $L^2(\mathbb{P})$ by assumption, concludes the proof.
\end{proof}

Geometric ergodicity is a well-studied concept in MCMC theory which ensures that the chain mixes at a fast rate; see \cite{Roberts2004}, Section 3.4., for a discussion of sufficient conditions, and Section 5.2.\@ for alternative sufficient conditions to obtain a CLT. Stronger results such as convergence almost surely or in probability could also be obtained using stronger conditions on the Markov chain; see for example Theorem 4 of \cite{Roberts2004} or Theorem 17.0.1 of \cite{Meyn1993}. Finally, all of the results aforementioned hold in the asymptotic setting where the number of MCMC samples $m\rightarrow \infty$. However, finite $m$ results could be obtained using concentration inequalities; see for example \cite{Paulin2015}.


\section{Additional Numerical Experiments}
\label{appendix:experiments}

In this appendix, we provide additional details on the numerical experiments in the paper including the Genz functions in Section \ref{appendix:genz}, the step function in Section \ref{appendix:step} and the Bayesian survey design experiment in Section \ref{appendix:survey_design}.

\begin{figure}[!tbh]
        \centering
        \includegraphics[width=0.7\linewidth,trim=0cm 0cm 0cm 0cm, clip]{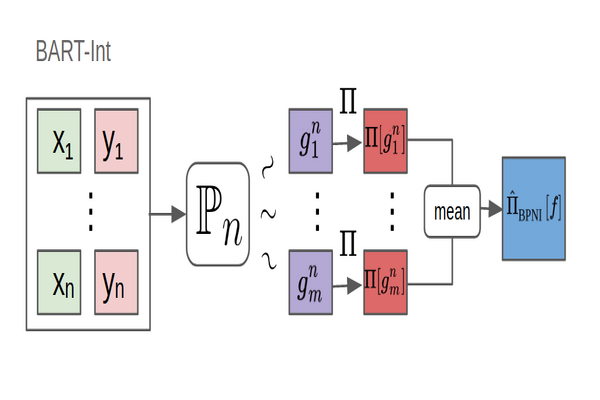}
        \caption{Bayesian Probabilistic Numerical Integration using Bayesian Additive Regression Trees (BART-Int). Here, $\Pi$ could either be known or replaced with an estimate $\hat{\Pi}$.}
        \vspace{-3mm}
        \label{figure:diagram}
\end{figure}

Figure~\ref{figure:diagram} provides a summary of the algorithm. We first observe some data pairs $\{(x_i,y_i)\}_{i=1}^n$ and then obtain the posterior on , which can be approximated by $m$ MCMC samples. We then integrate each of the samples and then take the mean to obtain an estimate of the integral $\hat{\Pi}_\text{BPNI}[f]$.

Our code relies on \verb+gpytorch+ \cite{gardner2018gpytorch} for kernel hyper-parameter (lengthscale) tuning for the GP, and \verb+dbarts+ \cite{dorie2018dbarts} as backend for implementing BART-Int. We use the Imperial College London High Performance Computing and the Department of Mathematics NextGen High Performance Computing servers to conduct our experiments. Our code is available on GitHub \url{https://github.com/ImperialCollegeLondon/BART-Int} and is subject to spontaneous maintenance.

\subsection{Genz Integrand Families}
\label{appendix:genz}

The Genz functions \cite{Genz} were taken from \url{http://www.sfu.ca/~ssurjano}, and are presented in Table \ref{tab:genz family}. They have two sets of parameters --- $d$ ``ineffective'' parameters $u$ and $d$ ``effective'' parameters $a$ which vary the level of difficulty. We use the default setting of $u = (0.5, \ldots, 0.5)^{\top}$ and scale $a$ suitably as the dimension increases to ensure numerical stability. Specifically, this is done by bounding the $L_1$-norm of $a$ so that numerical stability is obtained (see \cite{rescallingGenz} for details). As ground truth, we analytically compute the integrals for these Genz test functions, which are again given in Table \ref{tab:genz family}. We compare the performance of BART-Int for each function and make comparisons with two baselines: Monte Carlo integration (MI) and GP-BQ. Following the literature, we choose $\Pi$ to be the uniform distribution on $[0,1]^d$. 

For BART-Int, we used an MCMC sampler described in Appendix \ref{appendix:posterior} with a burn-in of $1000$ samples and took $5000$ samples afterwards. These were then thinned by keeping every $5$ samples. This led to $m=1000$. For the BART model, we used $T=50$ trees, and the pair $(\alpha, \beta) = (0.95, 2)$ for the terminating probability (see Appendix \ref{appendix:BART} for further details). We set $\sigma=0.1$ to calibrate its inverse-chi-squared prior \cite{dorie2018dbarts} due to our knowledge that there is no observation noise, but keep it to be non-zero to preserve the statistical properties of BART. For the rest of the hyper-parameters, we used the default setting from \verb+dbarts+. Note that we have applied very little tuning to the fitting of BART.

For GP-BQ, we used a prior mean $\mu(x)=0$ and the Mat\'ern kernel with smoothness $3/2$:
\begin{talign*}
k(x,y) = \left(1+\frac{\sqrt{3} \|x-y\|_2}{\rho}\right)\exp\left(-\frac{\sqrt{3}\|x-y\|_2}{\rho}\right),
\end{talign*}
where $\|\cdot\|_2$ is the Euclidean norm. The parameter $\rho$ is called the lengthscale, which was selected by maximising the marginal likelihood. To compute the kernel means, we used a MI estimate with $l=10^6$ randomly sampled points from $\Pi$. 

\begin{figure}[!tbh]
    \centering
    \begin{minipage}[t]{\textwidth}
        \centering
        \includegraphics[width=\linewidth,trim=0 0cm 0 0cm, clip]{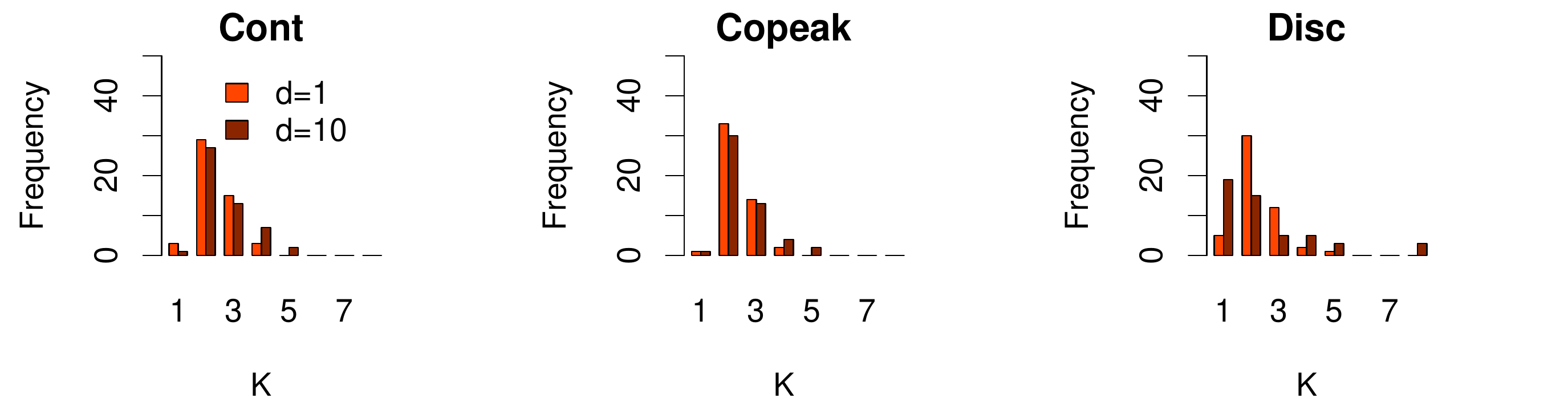}
        \includegraphics[width=\linewidth,trim=0 0cm 0 0cm, clip]{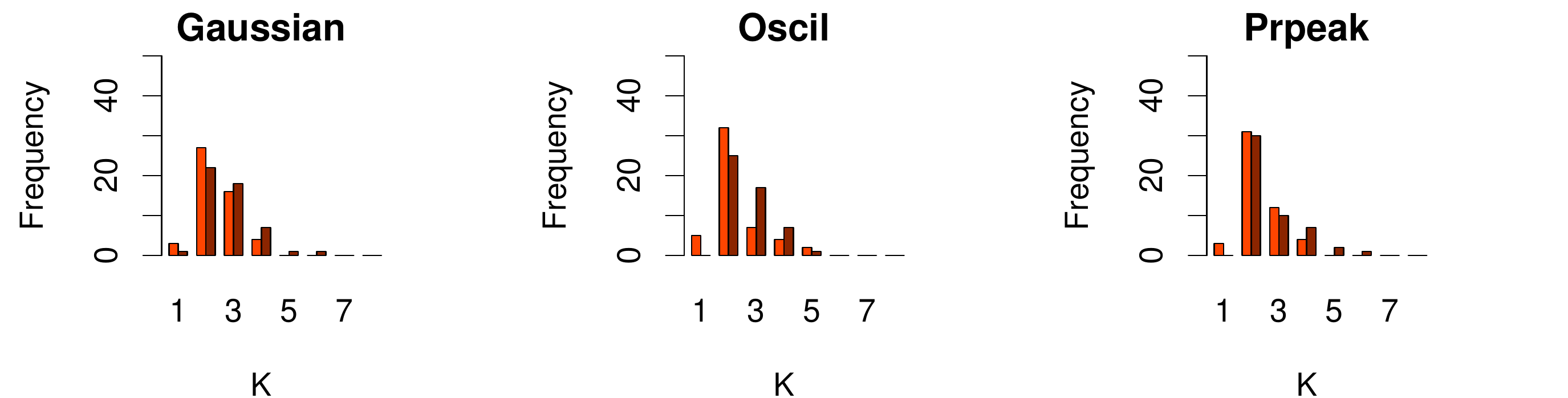}
        \caption{Histogram distribution of the number of leaf nodes, $K$ over $T=50$ trees for the Continuous, Corner Peak, Discontinuous, Gaussian, Oscillatory and Product Peak Genz function in dimensions $d=1,10$.}
        \label{fig:distribution_numberleaves_genz}
        \vspace{-3mm}
    \end{minipage}
\end{figure}

All of our results are presented in the main text. To complement these, we show the empirical distribution of the number of leaf nodes of the BART-Int method for each function in Figure \ref{fig:distribution_numberleaves_genz}. Recall that we chose the hyper-parameter values of $(\alpha,\beta)=(0.95,2)$ for the prior on trees, which guarantees that trees with $1, 2, 3, 4$ and $\geq$5 terminal nodes receive prior probability of $0.05$, $0.55$, $0.28$, $0.09$ and $0.03$ respectively.  As we can see, the posterior distribution of number of leaves per tree varies across target functions, demonstrating that BART is able to adapt to the target function.

For fixed targets, we see very little difference between the distribution for $d=1$ and $d=10$. This is sub-optimal since, as mentioned in \cite{rockova2017posterior}, the optimal number of leaves is $\textstyle \mathcal{O}\left(n^\frac{d}{2\alpha+d}\right)$ where $\alpha$ is the H\"older smoothness of the target function. For fixed $\alpha$, this suggests we should take a larger number of leaves in larger dimensions. This suggests further improvements in performance could be obtained by adapting the prior distribution as a function of $d$. On the other hand, the small number of leaves may also be seen as an advantage from a computational viewpoint.

\begin{sidewaysfigure}

\begin{landscape}
    \begin{minipage}[tbh]{\linewidth}
    \begin{adjustbox}{max width=\textwidth}
        \begin{tabular}[tbh]{| c | c | c | c |}
        \hline
          &  &  &  \\
          Family & Integrand  & Parameter & Integral \\
          &  &  &  \\
         \hline
          &  &  &  \\
          \shortstack{Continous \\ (cont)}
          & $\exp\left(-\sum_{i=1}^d a_i |x_i-u_i|\right)$ & $a_i=\frac{150}{d^3}$ & $\Pi_{i = 1}^{d} \frac{1}{a_i} \left( 2 - \exp(- a_i u_i) - \exp(a_i(u_i - 1)) \right)$\\
          &  &  &  \\
         \hline
          &  &  &  \\
          \shortstack{Corner Peak \\ (copeak)} & $\left(1 + \sum_{i=1}^{d}a_ix_i \right)^{-(d + 1)}$ & $a_i=\frac{600}{d^3}$ & $\left(\frac{1}{d! \Pi_{i = 1}^{d} a_i}\right)\sum\limits_{k = 0}^{d}\sum\limits_{\substack{I \subset \{1, \ldots, d \}\\ |I| = k}} (-1)^{k + d} \left( 1 + \sum\limits_{i = 1}^{d} a_i - \sum\limits_{j \in I} a_j \right)^{-1}$\\
          &  &  &  \\
         \hline
          &  &  &  \\
          \shortstack{Discontinuous \\ (disc)} & $\begin{cases}0, & \text{if $x_i > u_i$ for $i=1,\ldots,$ min(2, $d$)}\\
    \exp\left(\sum_{i=1}^da_ix_i\right), & \text{otherwise}\end{cases}$ & $a_i=\frac{10}{d^3}$ & $\prod_{i=1}^{\min(2, d)} \left[ \frac{1}{a_i} \left(e^{a_i \min(1, u_i)} - 1\right) \right] \prod_{i > 2} \left[ \frac{1}{a_i} \left(e^{a_i} - 1\right) \right]$\\
         &  &  &  \\
         \hline
         &  &  &  \\
          \shortstack{Gaussian Peak \\ (gaussian)} & $\exp\left(-\sum_{i=1}^da_i^2(x_i-u_i)^2\right)$ & $a_i=\frac{100}{d^2}$ & $\pi^{\frac{d}{2}} \Pi_{i = 1}^{d} \frac{1}{a_i}\left( \Phi(\sqrt{2} a_i (1 - u_i)) - \Phi(-\sqrt{2} a_i u_i) \right)$\\
          &  &  &  \\
         \hline
          &  &  &  \\
          \shortstack{Oscillatory \\ (oscil)}& $\cos(2\pi u_1 + \sum_{i = 1}^{d}a_ix_i)$ & $a_i=\frac{110}{d^{5/2}}$ & $\frac{1}{\Pi_{i = 1}^{d} a_i} \sum\limits_{k = 0}^{d} \sum\limits_{\substack{I \subset \{1, \ldots, d\} \\ |I| = k}} (-1)^{k} h_d(2\pi u_1 + \sum_{i = 1}^{d} a_i - \sum_{j \in I}a_j)$ \\
          &  &  & \shortstack{where $h_d(\cdot) = \begin{cases}\sin(\cdot) , &d \equiv 1 \mod 4 \\-\cos(\cdot) , & d \equiv 2 \mod 4 \\-\sin(\cdot) , & d \equiv 3 \mod 4 \\ \cos(\cdot) , & d \equiv 0 \mod 4 \end{cases}$}\\
          &  &  &  \\
         \hline
         &  &  &  \\
         \shortstack{Product Peak \\ (prpeak)} & $\Pi_{i=1}^d(a_i^{-2}+(x_i-u_i)^2)^{-1}$ & $a_i=\frac{600}{d^3}$ & $\Pi_{i = 1}^{d} a_i
        \left( 
            \arctan(a_i (1 - u_i)) - \arctan(-a_i u_i)
        \right)$\\
        &  &  &  \\
         \hline
        \end{tabular}
    \end{adjustbox}
    \captionof{table}{The Genz functions and their true integrals, defined with respect to a uniform measure on $[0,1]^d$. The default $u$-parameter is $u = (0.5, \ldots, 0.5)$ and the $a$-parameter is chosen such that the order of the integral values is comparable across dimension. Here, $\Phi(x)=\int_{-\infty}^x (2\pi)^{-1/2} \exp(-x^2/2) dx$ refers to the cumulative distribution function of a standard Gaussian random variable. A product over an empty set is 1 by convention (e.g.\ the second term in the integral of the discontinuous family when $d \leq 2$). }
    \label{tab:genz family}
    \end{minipage}
\end{landscape}
\end{sidewaysfigure}

\subsection{Step Function}
\label{appendix:step}

For the step function, we conducted experiments when integrating the function against either a uniform or a truncated Gaussian distribution. It is clear that with $\Pi$ being the uniform measure, $\Pi[f] = 0.5$ for all dimensions. When $\Pi$ is a multivariate Gaussian distribution with mean $x = (0.5, \ldots, 0.5)^\top$ and identity variance matrix truncated to $[0, 1]^d$, the integral is $\Pi[f] = 0.5$ (by symmetry). 

\begin{figure}[!tbh]
    \centering
    \begin{minipage}{\textwidth}
        \centering
        \includegraphics[width=\linewidth]{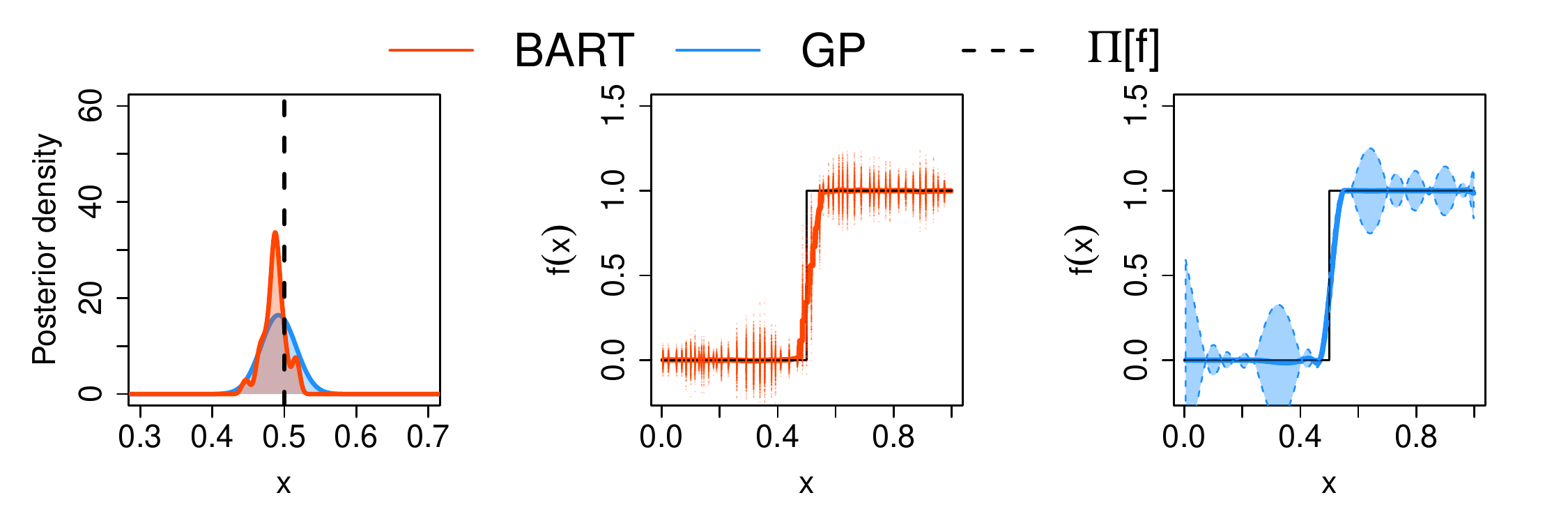}
        \caption{Integration of the step function against a uniform distribution over $[0,1]$ with BART-Int and GP-BQ with $n=20$ points. \textit{Left:} The posterior distribution on $\Pi[f]$. 
        \textit{Middle:} The BART posterior distribution on $f$. Posterior samples for BART are plotted as red points. \textit{Right:} The GP posterior distribution on $f$. The lines represent the posterior mean, the shaded areas give 95\% credible regions for the GP.}
        \vspace{-3mm}
        \label{fig:step_intro}
    \end{minipage}
\end{figure}

We first provide additional results when $\Pi$ is uniform. The performance of BART-Int in this case is presented in Section \ref{Performance}. Figure \ref{fig:step_intro} illustrates the posterior estimates for both the step function and its integral with BART-Int and GP-BQ. We can see that the posterior distribution of the integral for BART is more concentrated around the value of the true integral than the GP. We can also see that the GP has trouble estimating the discontinuity at $x=0.5$. Finally a disadvantage of BART-Int is that we see that uncertainty for both algorithms enlarge at areas where data is not observed, but for the end regions near 0 and 1 BART exhibits lower uncertainty due to its stepwise property. It is true that tree-based algorithms do not perform well for extrapolation tasks and this is quite evidently shown with the uncertainty intervals at the end points.

\begin{figure}[!tbh]
    \centering
    \begin{minipage}{\textwidth}
        \centering
        \includegraphics[width=1\linewidth,trim=0 2cm 0 2cm, clip]{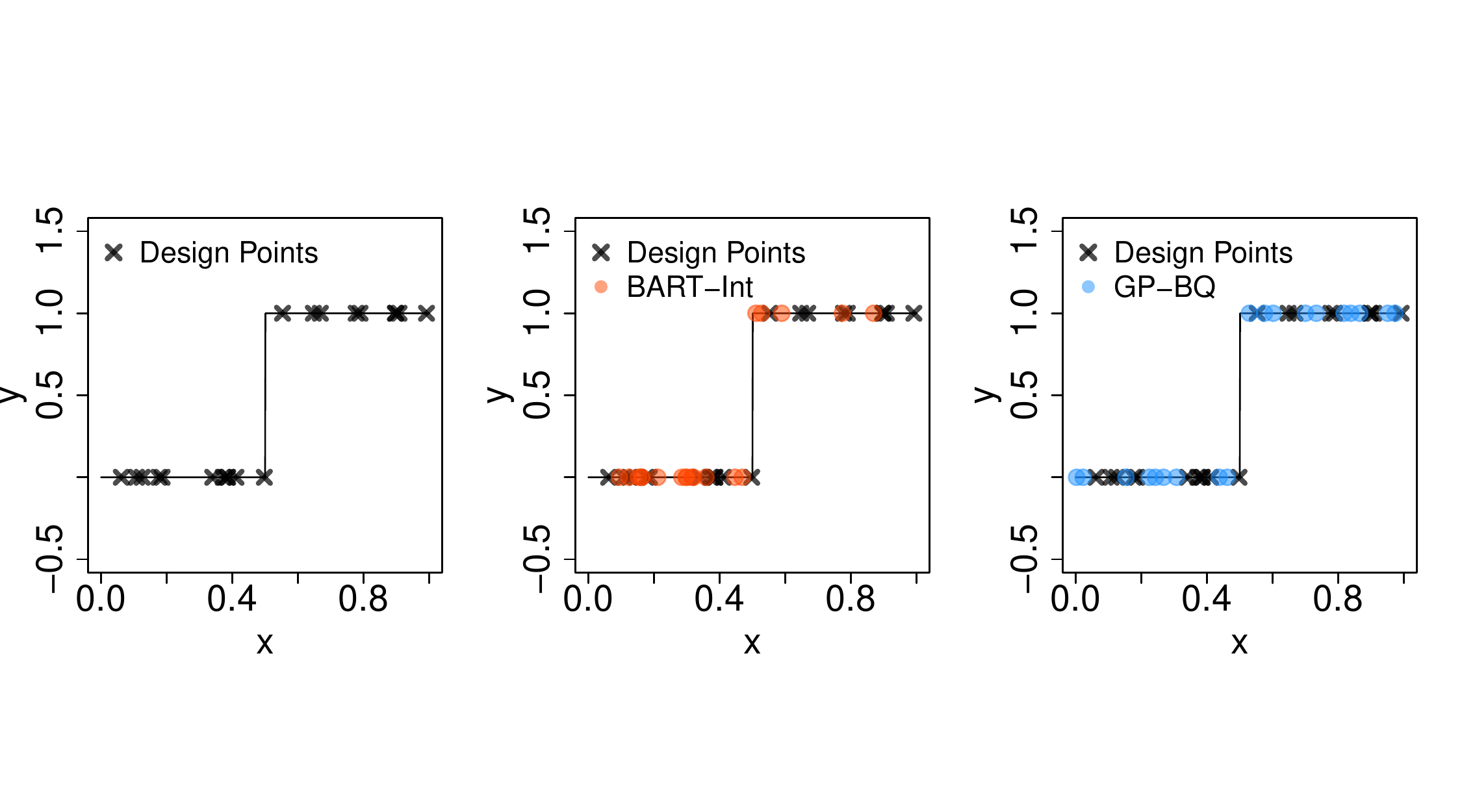}
        \caption{Illustration of adaptive selection of design points through Algorithm \ref{alg:SQ} on the step function integrated against a uniform measure over $[0,1]$ with $n_{\text{ini}}=20$ and $n_{\text{seq}}=20$.}
        \vspace{3mm}
        \label{figure:step_design}
    \end{minipage}
\end{figure}

Furthermore, Figure \ref{figure:step_design} illustrates the sequentially selected design points for each method. As we can see, both the BART and GP methods adaptively select points in areas not covered by the initial design points, and where the uncertainty about $f$ is hence greatest. 


\begin{figure}[!tbh]
        \centering
        \includegraphics[width=0.38\linewidth,trim=1cm 0cm 1cm 2cm, clip]{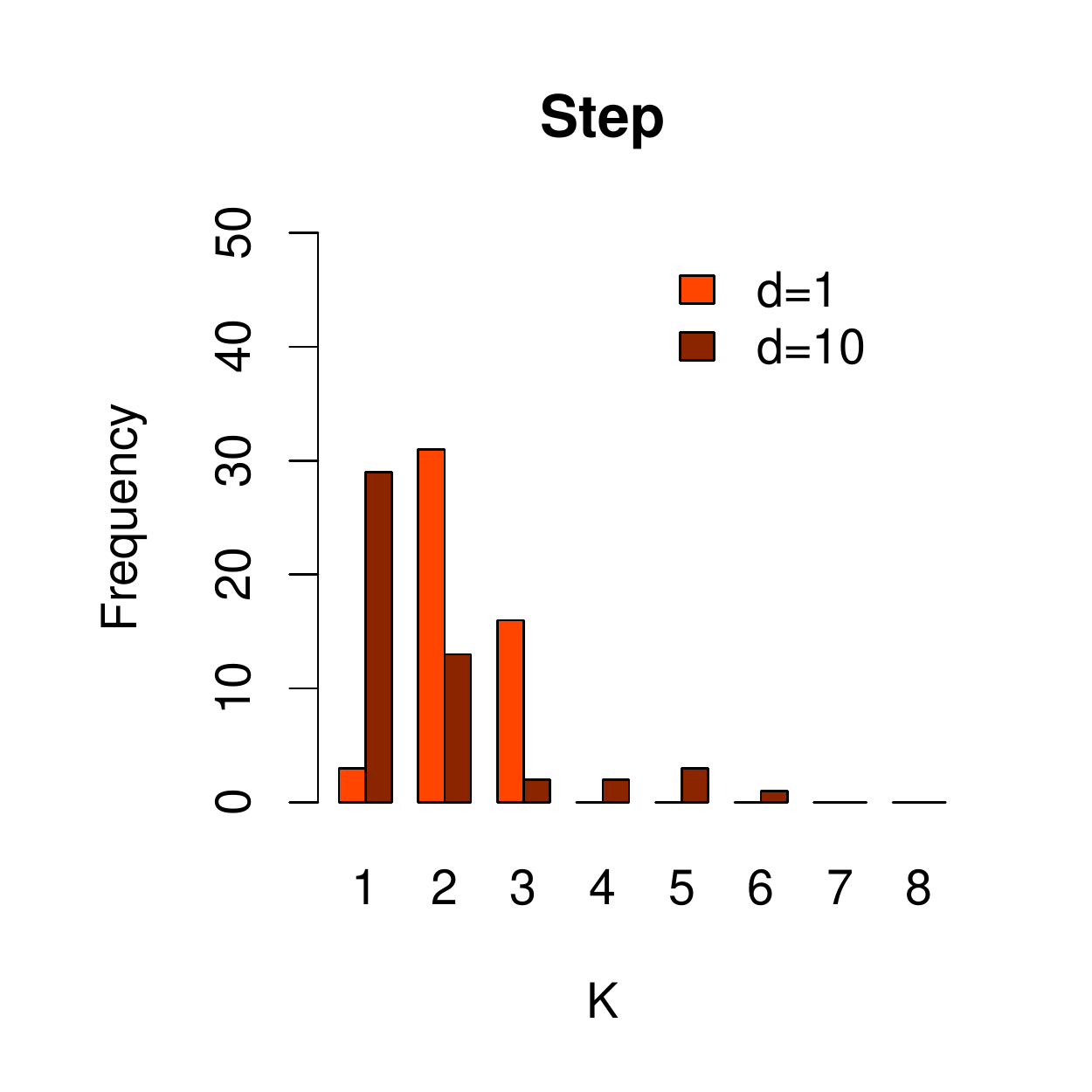}
        \caption{Histogram distribution of the number of leaf nodes, $K$ over $T=200$ trees for the step function in dimensions $d=1$ and $d=10$. }
        \vspace{-3mm}
        \label{figure:histogram_step_function}
\end{figure}

As another toy example, we ran BART-Int, GP-BQ and MI on the step function integrated against a truncated Gaussian measure. We started with $n_{\textrm{ini}} = 20$ design points and selected sequentially $n_{\textrm{seq}} = 20$ points according to the scheme introduced in Algorithm \ref{alg:SQ}. The experimental set-ups for BART-Int and GP-BQ were the same as in the previous experiment. The experiment was repeated with 20 sets of initial points that were sampled randomly and independently from $\Pi$. Over these runs, BART-Int achieved the smallest MAPE of $1.81$e-$02$, whereas MI gave $1.15$e-$01$ and GP-BQ yielded $2.83$e-$02$. 

\subsection{Bayesian Survey Design}
\label{appendix:survey_design}
To process the dataset in our experiments, we first randomly selected $n_\text{ini}=20$ points as our design points and another $10,000$ points as a candidate set. We then computed the logarithm of the income and created an indicator for each person: 1 if their log income is above 10 and 0 otherwise. 

All of the variables education, age, sex, own child, health insurance, marital status, employment and disability are categorical and we hence used a one-hot encoding. The education variable is an ordinal variable but we encoded it as a continuous variable for convenience. We then sampled $n_\text{seq}=200$ new points via sequential design using BART-Int and GP-BQ, and sampled randomly for MI. 

As a baseline ground truth, we used all $454,816$ points in the dataset and estimated the mean via MI. We also double-checked by using BART-Int with $10,020$ points from the design and candidate sets, which yielded very similar results.

We repeated this set-up $20$ times over different random initial points but the same candidate set. For GP-BQ, we set the lengthscale and $\sigma$ by maximising the marginal likelihood. For BART-Int, we mostly followed the default settings but used $T=50$ trees, $1000$ burn-in points, $5000$ posterior draws after burn-in and kept every $3$ draws from the posterior (thinning) so that $m=1666$. 

\end{document}